\pdfoutput=1
\documentclass[aps,prl,reprint,twocolumn,floatfix,%
superscriptaddress,longbibliography,nofootinbib]{revtex4-2}
\usepackage{tabularx}
\makeatletter
\def\hlinewd#1{%
\noalign{\ifnum0=`}\fi\hrule \@height #1
\futurelet\reserved@a\@xhline}
\makeatother
\usepackage{scrextend}

\usepackage{cellspace}
\usepackage{amsmath}
\usepackage{amssymb}
\usepackage{epsfig}
\usepackage{graphicx}
\usepackage{hyperref}
\usepackage{dcolumn}
\usepackage{slashed}
\usepackage{color}
\usepackage{rotating}
\usepackage[table,xcdraw,dvipsnames]{xcolor}
\usepackage[utf8]{inputenc}
\usepackage{colortbl}
\usepackage[normalem]{ulem}
\usepackage{mathrsfs}

\usepackage[vcentermath]{youngtab}
\usepackage{ytableau}

\usepackage{calligra}

\definecolor{nicered}{rgb}{0.7,0.1,0.1}
\definecolor{nicegreen}{rgb}{0.1,0.5,0.1}
\definecolor{red}{rgb}{1.0, 0, 0}

\newcommand{\SMlink}{\hyperref[S-sm]{Supplemental Material}}
\newcommand{\bdm}{\begin{displaymath}}
\newcommand{\edm}{\end{displaymath}}
\newcommand{\bea}{\begin{eqnarray}}
\newcommand{\eea}{\end{eqnarray}}

\definecolor{darkgreen}{rgb}{0.1,0.7,0.1}
\definecolor{nicered}{rgb}{0.7,0.1,0.1}
\definecolor{nicegreen}{rgb}{0.1,0.5,0.1}
\definecolor{red}{rgb}{1.0, 0, 0}
\definecolor{niceblue}{rgb}{0,0,0.8}
\hypersetup{colorlinks,citecolor= nicegreen,linkcolor= nicered,urlcolor=nicered}

\def\fig#1{{Fig.~\ref{#1}}}

\def\gsim{\raise0.3ex\hbox{$\;>$\kern-0.75em\raise-1.1ex\hbox{$\sim\;$}}}
\def\lsim{\raise0.3ex\hbox{$\;<$\kern-0.75em\raise-1.1ex\hbox{$\sim\;$}}}

\def\mb[#1]{\mathbf{#1}}

\definecolor{LightCyan}{rgb}{0.88,1,1}
\definecolor{piggypink}{rgb}{0.99, 0.87, 0.9}
\definecolor{applegreen}{rgb}{0.55, 0.71, 0.0}
\definecolor{darkpastelgreen}{rgb}{0.01, 0.75, 0.24}
\definecolor{green-yellow}{rgb}{0.68, 1.0, 0.18}

\newcommand{\beq}{\begin{equation}}
\newcommand{\eeq}{\end{equation}}
\newcommand{\beqa}{\begin{eqnarray}}
\newcommand{\eeqa}{\end{eqnarray}}

\newcommand{\Sec}[1]{ \medskip \noindent {\sl \bfseries #1}}

\usepackage[T1]{fontenc}
\usepackage{amsmath,amssymb,bm,graphicx,booktabs,array,microtype}
\usepackage{newunicodechar}
\usepackage{needspace}
\newunicodechar{−}{\ensuremath{-}}
\newunicodechar{ℓ}{\ensuremath{\ell}}

\newcommand{\lag}{\mathcal L}
\newcommand{\ham}{\mathcal H}
\newcommand{\Ima}{\operatorname{Im}}
\newcommand{\Rea}{\operatorname{Re}}
\newcommand{\dd}{\mathrm d}
\newcommand{\gev}{\mathrm{GeV}}
\newcommand{\mev}{\mathrm{MeV}}
\newcommand{\ecm}{e\,\mathrm{cm}}
\newcommand{\BR}{\mathcal B}
\newcommand{\Cs}{C_S}

\newcolumntype{L}[1]{>{\raggedright\arraybackslash}p{#1}}

\begin{document}

\title{Probing the CKM Origin of CP Violation with Paramagnetic EDMs \\ and Rare Kaon Decays}

\author{Giancarlo D'Ambrosio}
\email{gdambros@na.infn.it}
\affiliation{\small \it
INFN Sezione di Napoli, Complesso Universitario di Monte S.~Angelo,
Via Cintia Edificio 6, 80126 Napoli, Italy}
\author{Luca Di Luzio}
\email{luca.diluzio@pd.infn.it}
\affiliation{\small \it
INFN Sezione di Padova, Via Francesco Marzolo 8, 35131 Padova, Italy}

\begin{abstract}
In the Standard Model, paramagnetic electric dipole moments (EDMs) are dominated by kaon
exchange, which pairs the CP-violating short-distance transition $s\to d\,e^+e^-$ with a
CP-conserving weak kaon--nucleon coupling. Long-distance two-photon exchange, which dominates
$K_L\to e^+e^-$, gives an electron vertex about thirty-five times larger than the short-distance one and
could obscure this origin. We show that it does not: the two-photon vertex conserves CP and
must be paired with the CP-violating penguin component of the nucleon coupling, whose relative
size is $3\times10^{-4}$, so that electromagnetic corrections stay at the few-percent level.
Paramagnetic EDMs therefore measure $\mathrm{Im}(V_{ts}^*V_{td})$ and, together with the
CP-violating $S$ wave of $K_S\to\mu^+\mu^-$ and $K_L\to\pi^0\nu\bar\nu$, determine the same
imaginary coordinate of a kaon CKM plane with electrons, muons, and neutrinos, without
$B$-decay input. Lepton-specific new physics, or CP violation that bypasses kaon exchange,
would separate these determinations. Testing their common CKM origin calls for coordinated
progress in molecular EDM searches, rare-kaon experiments at LHCb, NA62, and KOTO, and
lattice determinations of weak nucleon matrix elements.
\end{abstract}

\maketitle

\Sec{Introduction.}
In the Standard Model (SM), the Cabibbo--Kobayashi--Maskawa (CKM) mechanism
\cite{Cabibbo:1963yz,Kobayashi:1973fv} relates the violation of
combined charge conjugation and parity (CP) in
processes separated by many orders of magnitude in energy.
Neutral kaons provided the first observation of CP violation
\cite{Christenson:1964fg} and remain central to testing its origin.
Today, rare kaon decays and CP violation
in kaon mixing test the same quark dynamics through complementary
amplitudes. The charged mode $K^+\to\pi^+\nu\bar\nu$, measured by
NA62~\cite{NA62:2026rwr}, constrains both the real and imaginary
$s\to d$ amplitudes, whereas
$K_L\to\pi^0\nu\bar\nu$~\cite{Littenberg:1989ix,Brod:2010hi}
and the CP-violating $S$ wave in
$K_S\to\mu^+\mu^-$~\cite{DAmbrosio:2017klp,Dery:2021mct}
isolate the imaginary component.
LHCb pursues the dimuon channel~\cite{DAmbrosio:2025mxa}, while
KOTO~\cite{KOTO:2024zbl} and the proposed KOTO II
experiment~\cite{KOTO:2025gvq} target the neutral neutrino mode.

At low energies, paramagnetic electric dipole moment (EDM) searches with atoms and
molecules offer an electron-sensitive addition to this programme.
Their CP-odd energy shifts probe the electron EDM and
scalar--pseudoscalar electron--nucleon interactions.
The ThO measurement by ACME~\cite{ACME:2018yjb}, the improved
HfF$^+$ result~\cite{Roussy:2022cmp}, and advances in ultracold YbF
interferometry~\cite{Jenkins:2026rko} and YbOH clock
transitions~\cite{Takahashi:2025lcb} motivate establishing the CKM
signal and its theoretical uncertainty.
Ema, Gao, and Pospelov~\cite{Ema:2022yra} recently identified a kaon-mediated
electron--nucleon interaction corresponding to an equivalent electron
EDM of order $10^{-35}\,\ecm$, about three orders of magnitude above
the earlier estimate~\cite{Pospelov:2013sca}.
Their contribution combines a CP-violating electroweak (EW)
$s\to d\,e^+e^-$ interaction with a CP-conserving weak
kaon--nucleon coupling. 
Unlike the CKM contributions to the neutron and diamagnetic EDMs, which are long-distance and
uncertain by an order of magnitude~\cite{Mannel:2012qk,Seng:2014lea,Yamanaka:2015ncb}, the paramagnetic one is
calculable in the SM and under control in chiral perturbation theory, with an uncertainty of about $30\%$~\cite{Ema:2022yra}.

\begin{figure}[!t]
 \centering
 \includegraphics[width=\columnwidth]{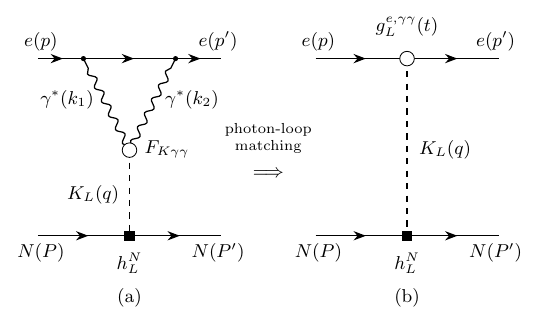}
 \caption{Two-photon $K_L$ exchange in elastic
 electron--nucleon scattering. The open blob $F_{K\gamma\gamma}$ in (a)
 is the CP-conserving $K_L\gamma^*\gamma^*$ form factor. The photon
 loop generates the pseudoscalar electron coupling
 $g_L^{e,\gamma\gamma}(t)$ in (b). The filled square is the CP-odd
 weak nucleon coupling $h_L^N$. The exchanged kaon carries spacelike
 momentum $t=q^2\simeq-\boldsymbol q^2$, close to zero in a molecule.
 The $K_S$ pole has the same topology, with the CP violation in the two-photon vertex.}
 \label{fig:KL-two-photon}
\end{figure}

That estimate keeps only the EW electron vertex, whereas the full vertex is 
the $K\to e^+e^-$ amplitude continued to small momentum transfer. 
Two-photon exchange is an established mechanism in rare kaon
decays~\cite{DAmbrosio:1986zin}: it dominates
$K_L\to\ell^+\ell^-$~\cite{Valencia:1997xe,DAmbrosio:1997eof,GomezDumm:1998gw,Isidori:2003ts,DAmbrosio:2023irq,Hoferichter:2023wiy}
($\ell=e,\mu$) and the total $K_S\to\ell^+\ell^-$
rate~\cite{Ecker:1991ru,DAmbrosio:2017klp,Dery:2021mct}. This raises a necessary question: could
a long-distance vertex overwhelm the EW one? In this Letter, we show that it does not. The
two-photon $K_L$ vertex (\fig{fig:KL-two-photon}) is about thirty-five times larger than the EW one,
but it conserves CP and must be paired with the small CP-violating component of the weak nucleon
coupling; in our hadronic estimate the correction is about $1\%$. The other electromagnetic (EM)
contributions, from the CP-violating two-photon $K_S$ vertex, CP-odd $\pi^0$ and $\eta$ couplings
to nucleons, and charged-pion loops, are also at the percent level, so the short-distance CKM
sensitivity survives.

We then place the paramagnetic EDM in the kaon CKM plane of
Ref.~\cite{Dery:2025pcx}, updated with current data, alongside the
CP-violating $S$ wave in $K_S\to\mu^+\mu^-$ and
$K_L\to\pi^0\nu\bar\nu$.
The plane constrains Wolfenstein combinations directly, so it needs no $B$-decay input
nor a $B$-dominated unitarity fit; agreement tests a common CKM origin
across electron, muon, and neutrino interactions.
Finally, we classify how new physics
can separate these determinations, discussing representative examples: CP violation
outside kaon exchange, such as an electron EDM, changes only the molecular signal;
lepton-dependent $s\to d$ interactions shift them
with the CKM phase, as for an electron-coupled neutral vector with quark minimal flavour violation (MFV), or with a new phase, as for a muon leptoquark.

\Sec{Kaon exchange and paramagnetic EDMs.}
The relevant semileptonic interaction is
\begin{equation}
 \lag_{eN}=\frac{G_F}{\sqrt2}\sum_{N=p,n}
          \Cs^N(\bar e i\gamma_5 e)(\bar N N),
 \label{eq:CSdefinition}
\end{equation}
with dimensionless coefficients $\Cs^N$.
For a nucleus of charge $Z$ and mass number $A$, the coherent
coupling is $\Cs^A=[Z\Cs^p+(A-Z)\Cs^n]/A$.
A molecular energy shift can be expressed as an equivalent electron
EDM, $d_{e,a}^{\rm eq}=d_e+\zeta_a\Cs^A$, where $\zeta_a$ is the
molecular conversion factor. For ThO,
$\zeta_{\rm ThO}\simeq1.5\times10^{-20}\,\ecm$
\cite{Denis:2016loy,Ema:2022yra}. The intrinsic CKM electron EDM,
$|d_e^{\rm CKM}|\sim6\times10^{-40}\,\ecm$
\cite{Yamaguchi:2020eub}, is negligible compared with the
semileptonic signal of order $10^{-35}\,\ecm$.
The quantum chromodynamics (QCD) vacuum angle also induces $\Cs$, through CP-odd pion and
$\eta$ couplings and two-photon exchange, with
$\Cs\simeq(0.016\text{--}0.03)\,\bar\theta$ for heavy nuclei
\cite{Flambaum:2019ejc,Mulder:2025esr}. It remains below $10\%$ of the
kaon-mediated CKM signal only for $|\bar\theta|\lesssim(2\text{--}4)\times10^{-15}$,
as in solutions of the strong CP problem that relax $\bar\theta$
dynamically.
A single paramagnetic EDM
measurement cannot separate the two sources.

Define the kaon vertices by
\begin{equation}
 \lag\supset\sum_{j=S,L}K_j
       [h_j^N\bar N N-g_j^e(t)\bar e i\gamma_5 e],\qquad t=q^2.
 \label{eq:vertices-main}
\end{equation}
Here $q=p'-p$, where $p,p'$ are the incoming and
outgoing electron momenta. Kaon exchange then gives
\begin{equation}
 \Cs^N(t)=-\frac{\sqrt2}{G_F}
       \sum_{j=S,L}\frac{h_j^N g_j^e(t)}{m_K^2-t}.
 \label{eq:KLmain}
\end{equation}
For a nearly static nucleus of mass $M_A$, recoil gives
$|q^0|\simeq|\boldsymbol q|^2/(2M_A)\ll|\boldsymbol q|$, so
$t\simeq-|\boldsymbol q|^2$.
At the thorium Coulomb scale
$|\boldsymbol q|\sim Z\alpha m_e\simeq0.34\,\mev$,
$M_{\rm Th}\simeq216\,\gev$ gives a recoil
$|q^0|\simeq0.26\,\mathrm{eV}$ and $|t|/m_K^2\simeq5\times10^{-7}$.
Hence, the kaon propagator is accurately approximated by its
zero-transfer value and we evaluate the local coefficient $\Cs^N(0)$,
while kaon decays probe the timelike point $t=m_K^2$.

Here $K_S$ and $K_L$ denote the CP-even and CP-odd states. Indirect
CP violation does not alter this assignment for the molecule: at
$|t|\ll m_K^2$ the $\varepsilon_K$ admixtures of the two mass
eigenstates cancel between the $K_S$ and $K_L$ poles (Sec.~\ref{S-app:matching} of the
Supplemental Material~\cite{SupplementalMaterial}).
Since $\bar e i\gamma_5 e$ is CP odd and $\bar N N$ CP even, the EDM requires CP
violation at the electron vertex for $K_S$ exchange and at the
nucleon vertex for $K_L$ exchange. A large CP-conserving $K_L$
electron amplitude contributes only when multiplied by $h_L^N$.
We write the electron source as $E=\lambda_uD_u+\lambda_tD_t$, with
$g_S^e=4F_\pi m_e\Ima E$ and $g_L^e=4F_\pi m_e\Rea E$, and the weak
nucleon coefficient as $c_N=\lambda_u^*c_{Nu}+\lambda_t^*c_{Nt}$, with
$h_S^N=\Rea c_N/F_\pi$ and $h_L^N=\Ima c_N/F_\pi$. The coefficients
$D_u$, $D_t$, $c_{Nu}$ and $c_{Nt}$ are real: all weak phases are carried
by the CKM factors $\lambda_i=V_{is}^*V_{id}$. The $K_S$ and $K_L$ poles then combine
at $t\simeq0$ into
\begin{equation}
 h_S^Ng_S^e+h_L^Ng_L^e=4m_e\,J\,(D_tc_{Nu}-D_uc_{Nt}),
 \label{eq:invariant-main}
\end{equation}
with $J=\Ima(\lambda_t\lambda_u^*)$.
The EW mechanism of Ref.~\cite{Ema:2022yra} is the
$D_tc_{Nu}$ term. The two-photon $K_L$ contribution belongs to
$D_uc_{Nt}$ and therefore requires the CP-violating penguin part
$c_{Nt}$ of the nucleon coupling.

Conventions, CP assignments, and the weak nucleon
matching are collected in Sec.~\ref{S-app:matching}.

\Sec{The two-photon correction.}
In the quark phase convention with real $V_{us}^*V_{ud}$, the
leading EW electron coupling is
\begin{equation}
 g_S^{e,\rm EW}=-
 \frac{\sqrt2G_F\alpha_{\rm em}(m_Z)F_\pi m_e}{\pi s_W^2}
 \,\Ima(V_{ts}^*V_{td})Y(x_t).
 \label{eq:gSEW-main}
\end{equation}
Here $s_W$ is the sine of the weak mixing angle,
$x_t=m_t^2/m_W^2$, and $Y(x_t)$ is the charged-lepton Inami--Lim
function. The electron mass reflects the required chirality flip.
The factor $F_\pi=92.07\,\mev$ comes from the kaon axial-current
matrix element, using $f_K=\sqrt2F_\pi$ at leading SU(3) chiral
order. The same decay constant enters inversely in $h_S^N$ and
cancels from the leading $\Cs^N$; see Sec.~\ref{S-app:matching}.
For the leading-order (LO) matching benchmark, with
$\Ima(V_{ts}^*V_{td})=1.42\times10^{-4}$,
$g_S^{e,\rm EW}=-1.13\times10^{-15}$.

For two-photon exchange we normalize
$\lag_{K\gamma\gamma}=-c_\gamma K_LF_{\mu\nu}\widetilde F^{\mu\nu}/4$,
where $\widetilde F^{\mu\nu}=\epsilon^{\mu\nu\rho\sigma}F_{\rho\sigma}/2$.
The measured diphoton width~\cite{ParticleDataGroup:2024cfk} gives $|c_\gamma|=[64\pi\Gamma(K_L\to\gamma\gamma)/m_K^3]^{1/2}=3.39\times10^{-9}\,\gev^{-1}$.
We retain this magnitude at the soft kaon point, without additional
derivative-mixing suppression, and include a vector-meson-dominance form factor for the virtual photons
\cite{DAmbrosio:1997eof,Isidori:2003ts}.

The photon-loop projection illustrated in
Fig.~\ref{fig:KL-two-photon} gives
\begin{align}
 g_L^{e,\gamma\gamma}(0)
   &=\frac{\alpha(0)m_ec_\gamma}{2\pi}\,\mathcal A(0),\nonumber\\
 \mathcal A(0)
   &\simeq3\ln\frac{m_e}{M_V}-\frac{11}{4}-3a_V=-19.9,
 \label{eq:mainnumbers}
\end{align}
with the vector-meson scale fixed to the $\rho$ mass,
$M_V=0.775\,\gev$ and slope $a_V=-1.611$~\cite{Isidori:2003ts}.
Thus $|g_L^{e,\gamma\gamma}(0)|=4.00\times10^{-14}$, about sixteen times its EW $K_L$ counterpart and thirty-five times $|g_S^{e,\rm EW}|$.
Sec.~\ref{S-app:loop} gives the soft-source assumptions,
the loop calculation, and its matching onto $\Cs^N$.

The nucleon vertex requires independent CP-odd baryon matrix
elements. We retain the real weak baryon couplings fitted to
hyperon decays in Ref.~\cite{Ema:2022yra} and estimate their
imaginary parts by matching the gluonic
penguin operator
\begin{equation}
 Q_6=(\bar d_\alpha\gamma_\mu(1-\gamma_5)s_\beta)
 \sum_{q=u,d,s}
 (\bar q_\beta\gamma^\mu(1+\gamma_5)q_\alpha)
 \label{eq:Q6operator}
\end{equation}
onto the same chiral structures, where $\alpha,\beta$ are colour
indices. $Q_6$ dominates the CP-violating part because its Fierz form is a product of scalar densities, whose
chirally enhanced baryon matrix elements we estimate using factorized and bag-model
contributions~\cite{Tandean:2002vy,He:2025wxn}.
For ${}^{232}$Th, the coherent average
$h_j^{\rm Th}=(90h_j^p+142h_j^n)/232$ gives
$h_S^{\rm Th}=-8.05\times10^{-7}$ and
$h_L^{\rm Th}/h_S^{\rm Th}=3.18\times10^{-4}$.
This is comparable to
$|\Ima A_0/\Rea A_0|\simeq2\times10^{-4}$ in the isospin-zero
$K\to\pi\pi$ amplitude $A_0$ (strong phase removed), whose
imaginary part is also dominated by $Q_6$~\cite{RBC:2020kdj}; both are of order
$\Ima\lambda_t/\lambda_u\simeq6.5\times10^{-4}$ times a hadronic factor of order one.
Eq.~\eqref{eq:KLmain} therefore yields
\begin{align}
 \frac{|\Delta\Cs^{\rm Th,L,\gamma\gamma}|}
      {|C_{S,\rm EW,LO}^{\rm Th}|}
 &\simeq
 \left|\frac{h_L^{\rm Th}}{h_S^{\rm Th}}\right|
 \left|\frac{g_L^{e,\gamma\gamma}(0)}{g_S^{e,\rm EW}}\right|
 \nonumber\\
 &=(3.18\times10^{-4})(35.5)=1.1\%.
 \label{eq:firstresult}
\end{align}
Relative to the value $|C_{S,\rm EW}^{\rm Th}|=6.9\times10^{-16}$ of
Ref.~\cite{Ema:2022yra}, the correction is $0.73\%$. It is
conservative in two respects: we impose no derivative-mixing
suppression at $t=0$, and the local evaluation ignores the fall of
the photon loop at nuclear momenta,
$|\mathcal A(-Q^2)/\mathcal A(0)|\simeq0.58$ and $0.26$ at
$Q\simeq5$ and $15\,\mev$ (Sec.~\ref{S-app:loop}). The remaining EM terms are of
similar size. The $K_S$ pole also contributes, through a CP-violating two-photon coupling with weak phase $\rho_\gamma=g_S^{e,\gamma\gamma}/g_L^{e,\gamma\gamma}$; only the sum $h_L^{\rm Th}/h_S^{\rm Th}+\rho_\gamma$ is independent of phase conventions. Kaon mixing cancels between the poles (see above), and the $Q_6$ penguin cancels in $K_L\to\gamma\gamma$~\cite{Gerard:2005yk}. Hence $\rho_\gamma\simeq2\times10^{-5}$ and the kaon-pole term becomes $0.76\%$; it stays
at the $1\%$ level even if the soft photon vertex has a larger penguin share (Sec.~\ref{S-app:loop}). We estimate that CP-odd
$\pi^0,\eta$ couplings and charged-pion loops add
$2.4\%$~\cite{Yamanaka:2015ncb,Mulder:2025esr} (Sec.~\ref{S-app:loop}). EM
effects are therefore at the few-percent level, well below the
$30\%$ uncertainty of the CP-conserving weak kaon--nucleon couplings
that normalize the leading term~\cite{Ema:2022yra}.

\begin{figure*}[!t]
 \centering
 \includegraphics[width=\textwidth]{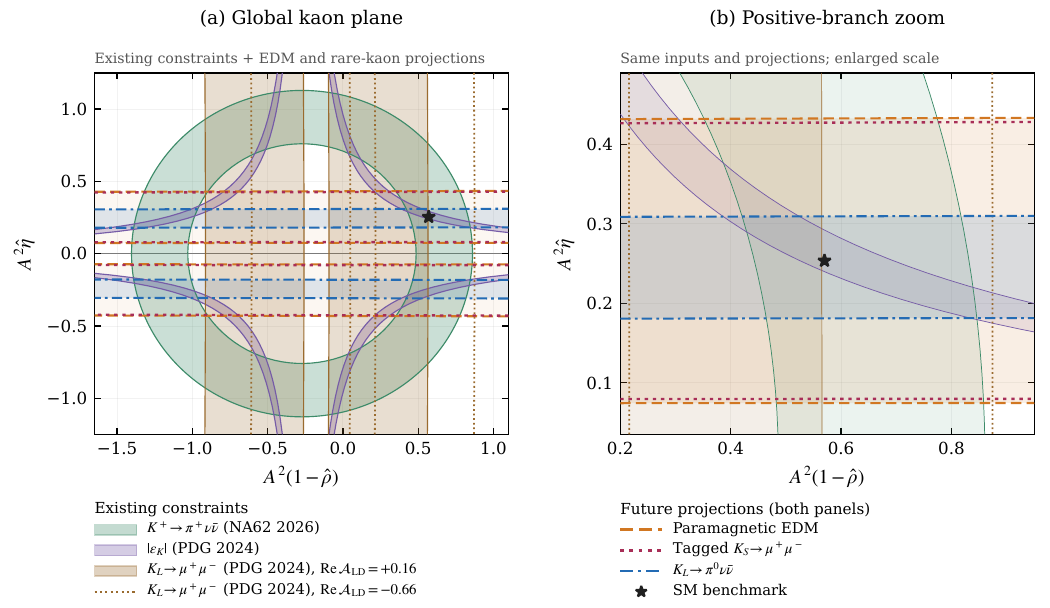}
 \caption{Paramagnetic EDMs in the kaon CKM
 plane of Ref.~\cite{Dery:2025pcx}. Both panels show the existing
 NA62 2016--2024~\cite{NA62:2026rwr} and
 $|\varepsilon_K|$ constraints (solid-bordered green and
 violet regions), together with future projections for the EDM
 (orange dashed), tagged $K_S\to\mu^+\mu^-$ (red dotted), and
 $K_L\to\pi^0\nu\bar\nu$ (blue dash-dotted boundaries with
 blue shading). Each pair of broken boundaries, except the brown dotted ones, encloses a projected
 interval. Brown: measured $K_L\to\mu^+\mu^-$ rate for the two possible signs of the
 $K_L\to\gamma\gamma$ amplitude (shaded and dotted)~\cite{Hoferichter:2023wiy,Dery:2025pcx}; each gives
 two bands. Panel (a) retains both signs of $A^2\hat\eta$, since
 decay rates determine its magnitude; panel (b) enlarges the positive
 branch. The black star is a chosen SM
 benchmark, $(A^2(1-\hat\rho),A^2\hat\eta)=(0.57,0.2535)$, used
 to centre the future projections. The individual nominal $95\%$ projection
 bands assume $20\%$ experimental and $30\%$ weak-nucleon errors
 for the EDM, $35\%$ muon-amplitude sensitivity~\cite{DAmbrosio:2025mxa}, and $25\%$
 neutrino-rate precision~\cite{KOTO:2025gvq}.}
 \label{fig:kaon-plane}
\end{figure*}

\Sec{Paramagnetic EDMs in the kaon CKM plane.}
We employ the Wolfenstein combinations $A^2(1-\rho)$ and
$A^2\eta$~\cite{Wolfenstein:1983yz} as the LO
coordinates of the kaon CKM plane.
Following Ref.~\cite{Dery:2025pcx}, we define their exact
kaon-triangle counterparts by
\begin{equation}
 A^2(1-\hat\rho+i\hat\eta)
 =\frac{1}{\lambda^4}
   \left(\frac{V_{ts}^*V_{td}}{V_{cs}^*V_{cd}}\right)^* .
 \label{eq:kaon-plane}
\end{equation}
Here $\lambda\simeq|V_{us}|$ is determined from kaon decays;
$\hat\rho,\hat\eta$ agree with the familiar Wolfenstein parameters
at LO. The two axis combinations are determined
directly, without separately fixing $A\simeq|V_{cb}|/\lambda^2$
from $B$ physics. At LO,
$\Ima(V_{ts}^*V_{td})\simeq\lambda^5 A^2\hat\eta$.
The exact conversion, observable formulae, inputs, and uncertainty treatment used for the figure are given in Sec.~\ref{S-app:kaonplane}.

The established constraints supply complementary geometry: the
$K^+\to\pi^+\nu\bar\nu$ rate gives a nearly circular band,
shifted to negative $A^2(1-\hat\rho)$ by its charm contribution, and
$\varepsilon_K$ a hyperbolic band. The dispersive $K_L\to\mu^+\mu^-$
amplitude adds real-axis information (Fig.~\ref{fig:kaon-plane}, both signs), with long-distance and sign
ambiguities~\cite{GomezDumm:1998gw,Hoferichter:2023wiy,Dery:2025pcx}. Independent CP
probes can test the common imaginary coordinate.
Three such determinations involve different leptons. At leading
Wolfenstein order,
\begin{equation}
 \begin{aligned}
 C_{S,\rm EW}^{A}&\propto A^2\hat\eta\,Y(x_t),\\
 \BR_S^{(0)}&\propto[A^2\hat\eta\,Y(x_t)]^2,\\
 \BR(K_L\to\pi^0\nu\bar\nu)&\propto[A^2\hat\eta\,X(x_t)]^2.
 \end{aligned}
 \label{eq:plane-CP-probes}
\end{equation}
Here $X(x_t)$ is the neutrino short-distance function and
$\BR_S^{(0)}$ is the branching ratio of the CP-violating,
spin-singlet $S$ wave in $K_S\to\mu^+\mu^-$.
Its extraction requires tagged neutral-kaon interference, since
the total $K_S$ dimuon rate is dominated by the CP-conserving
$P$ wave~\cite{DAmbrosio:2017klp,Dery:2021mct,Dery:2022yqc}.
In this interference the small $K_S$ amplitude multiplies the large $K_L\to\mu^+\mu^-$ amplitude, which is fixed by measured rates, so the long-distance origin of the latter does not bias the extraction.

Fig.~\ref{fig:kaon-plane} adds the paramagnetic EDM determination to
this picture. The three CP observables give overlapping, nearly
horizontal bands in the SM; the EDM adds an electron-sensitive
test, with a different hadronic normalization. Rate measurements alone leave the sign of
the imaginary coordinate ambiguous; both branches are retained
in Fig.~\ref{fig:kaon-plane}.
The molecular EDM signal is instead linear in
$\Ima(V_{ts}^*V_{td})$: once the
overall sign of the CP-conserving weak nucleon couplings is fixed,
for example by lattice QCD, its sign selects the branch.

\Sec{New-physics tests with molecules and kaons.}
Departures from the common CKM interpretation can arise from
flavour-diagonal CP violation generating an electron EDM or a
local electron--nucleon interaction, as well as from new
strangeness-changing amplitudes. We illustrate both classes,
retaining the SM weak nucleon couplings in the kaon-exchange
examples below.

\emph{Direct EW contributions to the molecular EDM.}
New CP phases in the Higgs or EW sector can generate
$d_e$ without requiring a new $s\to d$ coupling. For example,
a CP-odd Higgs--top interaction
$\lag\supset-(m_t/v)\widetilde\kappa_t h\bar t i\gamma_5t$,
with $v=246\,\gev$ and the SM electron Yukawa coupling, induces
$d_e\simeq9\times10^{-27}\widetilde\kappa_t\,\ecm$
through a two-loop Barr--Zee diagram~\cite{Brod:2013cka}.
Thus $\widetilde\kappa_t=10^{-4}$ gives
$d_e\simeq9\times10^{-31}\,\ecm$, almost five orders of
magnitude above the CKM-induced molecular EDM signal and below the
HfF$^+$ limit $4.1\times10^{-30}\,\ecm$
when interpreted with other EDM sources absent~\cite{Roussy:2022cmp}.
Even three-loop contributions from new EW multiplets
can exceed the CKM signal~\cite{Banno:2024apv}; a benchmark is
given in Sec.~\ref{S-app:NP}.
Independent local contributions to $\Cs^A$ provide another
possibility, entering as
$\Delta d_{e,a}^{\rm eq}=\Delta d_e+\zeta_a\Delta\Cs^A$
\cite{Ardu:2025rqy}. Such effects can produce a molecular EDM
discrepancy while both rare-kaon measurements remain compatible
with the SM.

\emph{An electron-specific interaction with only the CKM phase.}
Consider a heavy neutral vector $Z'$ with an $s\to d$ coupling
$g_qV_{ts}^*V_{td}$ to left-handed quarks and a coupling $g_e$ to
right-handed electrons. Both $g_q$ and $g_e$ are real.
A realization with quark MFV
also couples it to top and bottom quarks with strength approximately
$g_q$; quark MFV does not require lepton universality
\cite{DAmbrosio:2002vsn}.
At tree level it replaces $Y(x_t)$ in the electron amplitude by
$Y(x_t)+\delta Y_e$, where
\begin{equation}
 \delta Y_e=+0.113\left(\frac{g_q}{0.5}\right)g_e
                  \left(\frac{5\,\mathrm{TeV}}{M_{Z'}}\right)^2.
 \label{eq:Zbenchmark-main}
\end{equation}
The muon and neutrino amplitudes are unchanged at this order.
For $M_{Z'}=5\,\mathrm{TeV}$, $g_q=0.5$, and $g_e=1$,
the molecular EDM coefficient and its inferred $A^2\hat\eta$ increase by $12\%$; reversing $g_e$ gives a reduction.
It satisfies the LEP
right-handed electron contact limit, $M_{Z'}/|g_e|>2.2\,\mathrm{TeV}$
\cite{ALEPH:2013dgf}. The associated four-quark interaction
increases the top-quark contribution to neutral-meson mixing by
about $9\%$, providing a correlated test through $\varepsilon_K$
and $B$ mixing. With $g_q=0.5$ and $M_{Z'}=5\,$TeV fixed, the electron contact limit alone allows the molecular EDM coefficient, and the inferred $A^2\hat\eta$, to shift by up to $27\%$ (Sec.~\ref{S-app:NP}).

Larger effects are possible when the $\bar sd$--electron interaction comes without tree-level
four-quark and four-electron terms, as for a leptoquark coupled to electrons.
Kaon decays to electrons constrain such $\bar sd$--electron couplings directly.
The KLOE limit $\BR(K_S\to e^+e^-)<9\times10^{-9}$
\cite{KLOE:2008acb} implies
$|d_{e,\rm ThO}^{\text{eq},K_S}|\lesssim3.4\times10^{-31}\,\ecm$,
about 30 times below the current ThO limit, but only for a purely
axial or pseudoscalar, maximally CP-violating coupling.
A current with a vector component, as for the $Z'$, is constrained far more strongly by
$K_L\to\pi^0e^+e^-$, which limits the kaon-mediated EDM signal to
$|d_{e,\rm ThO}^{\text{eq}}|\lesssim10^{-34}\,\ecm$, about ten times the SM
\cite{KTeV:2004piee,Mescia:2006jd} (Sec.~\ref{S-app:NP}).

\emph{A muon-specific leptoquark with an additional weak phase.}
A scalar leptoquark $\widetilde S_1$ in the representation
$(\bar{\boldsymbol3},\boldsymbol1,4/3)$ of
$SU(3)_c\times SU(2)_L\times U(1)_Y$ can couple to $d_R,s_R$
and $\mu_R$, generating
$\ham\supset c_R^\mu(\bar s_R\gamma_\alpha d_R)
(\bar\mu_R\gamma^\alpha\mu_R)+\mathrm{h.c.}$
\cite{Dorsner:2016wpm,Dery:2021vql}.
It changes the muon determination without a tree-level electron
or neutrino counterpart. In the convention with real $V_{us}^*V_{ud}$,
the normalization used here gives
\begin{equation}
 \BR_S^{(0)}\simeq1.9\times10^{-13}
 \left|1-\frac{\Ima c_R^\mu}
                    {2.4\times10^{-11}\,\gev^{-2}}\right|^2.
 \label{eq:LQbenchmark-main}
\end{equation}
A preliminary LHCb analysis reports a
$K^0\to\mu^+\mu^-$ signal with $4.5\sigma$ significance.
Combining this analysis with earlier searches gives
$\BR(K_S\to\mu^+\mu^-)<2.7\times10^{-11}$
\cite{LHCb:2026ksmumu},
leaving room for an $S$-wave rate of order $10^2$
times its SM value.
For a concrete example, $M_{\widetilde S_1}=3\,\mathrm{TeV}$
and $c_R^\mu=-i\,1.2\times10^{-10}\,\gev^{-2}$ require leptoquark
couplings of about $0.046$ to both $d_R\mu_R$ and $s_R\mu_R$ and give
$\BR_S^{(0)}\simeq6.8\times10^{-12}$, about $36$ times the SM
rate. The purely imaginary coefficient leaves the leading
CP-conserving $K_L\to\mu^+\mu^-$ amplitude unchanged; its square
is real, suppressing the new contribution to $\varepsilon_K$.
The resulting kaon mass-splitting correction is about $4\times10^{-4}$.
The accompanying $K_L\to\pi^0\mu^+\mu^-$ rate also
satisfies its current bound; the matching and kaon constraints
are given in Sec.~\ref{S-app:NP}.

\emph{Separating the three determinations.}
Interpreting each observable with its SM formula yields an
apparent value of $A^2\hat\eta$. The neutral-vector benchmark
shifts the paramagnetic EDM determination, whereas the leptoquark
shifts the CP-violating $K_S\to\mu^+\mu^-$ determination;
$K_L\to\pi^0\nu\bar\nu$ is unchanged at tree level in both. More generally, for MFV operators and SM predictions known to $10\%$, the paramagnetic EDM probes scales near $5\,$TeV, comparable to $K_L\to\pi^0\nu\bar\nu$ and the CP-violating $K_S\to\mu^+\mu^-$ $S$ wave, and it is the most sensitive probe of operators coupled only to electrons (Fig.~\ref{S-fig:MFV}).
An isolated paramagnetic EDM discrepancy, however, cannot distinguish
a modified electron--kaon interaction from a direct electron EDM
or a local electron--nucleon interaction. Measurements in several
EDM systems with sufficiently different sensitivities can help
separate $d_e$ from $\Cs^A$~\cite{Fleig:2018bsf,Degenkolb:2024eve}.
Lepton-doublet interactions, scalar operators, and light mediators
give further patterns (Sec.~\ref{S-app:NP}). Measuring
all three channels
tests their common CKM interpretation
and helps identify departures from it.

\Sec{Conclusions and outlook.}
Our main result is that long-distance two-photon exchange, which
dominates $K_L\to e^+e^-$, changes the CKM paramagnetic signal only
at the percent level, because the large CP-conserving
$K_L\gamma\gamma$ vertex must be paired with the small CP-violating
penguin component of the weak kaon--nucleon coupling; together with
the other EM terms, the correction stays at the
few-percent level. Paramagnetic EDMs therefore measure the
short-distance quantity $\Ima(V_{ts}^*V_{td})$ and supply the electron leg of a kaon
CKM test that requires no $B$-physics input.
The main theoretical priority is
the weak nucleon matrix elements, including their sign and
CP-violating parts, from hyperon data, chiral methods, and lattice
QCD.

The kaon CKM plane of
Ref.~\cite{Dery:2025pcx} gives a common target for NA62, LHCb, KOTO, and
molecular EDM measurements. A phenomenological study of a tagged analysis with an LHCb-like
detector at the high-luminosity LHC finds sensitivity near
$35\%$ of the SM CP-violating dimuon amplitude in its optimistic
scenario~\cite{DAmbrosio:2025mxa}, and KOTO II targets
$25\%$ precision on the neutral-neutrino rate, i.e.\ $12.5\%$ on its
amplitude~\cite{KOTO:2025gvq}. A $20\%$ measurement of the CKM
paramagnetic signal requires an equivalent-EDM sensitivity of a few
$10^{-36}\,\ecm$, a long-term target motivated by progress in
molecular preparation, coherence, and statistical
reach~\cite{Vutha:2017pej,Alarcon:2022ero,EuropwanEDMprojects:2025okn}. Reaching it
would turn paramagnetic EDMs from null tests into precision probes of
the CKM origin of CP violation, while larger new-physics effects can
make the comparison informative earlier.

\begin{acknowledgments}
\Sec{Acknowledgments.}
We thank Riccardo Barbieri, Gino Isidori and Paride Paradisi for very relevant discussions.
GD acknowledges support
from the INFN project ENP.
The work of LDL is supported by the Italian Ministry of University and Research (MUR) via the FIS2 Consolidator Grant project FIS-2023-02106 -- QAXION (CUP: I53C25001880001). We acknowledge the use of Claude
(Anthropic) for assistance with calculations and manuscript preparation; the authors
take full responsibility for the results and conclusions.
\end{acknowledgments}

\bibliographystyle{apsrev4-1.bst}
\bibliography{Bibliography_PRL_with_SM}

\clearpage
\appendix
\renewcommand{\appendixname}{}
\setcounter{section}{0}\setcounter{subsection}{0}
\setcounter{equation}{0}\setcounter{table}{0}\setcounter{figure}{0}
\counterwithout{equation}{section}
\renewcommand{\thesection}{S\arabic{section}}
\renewcommand{\theequation}{S\arabic{equation}}
\renewcommand{\thetable}{S\arabic{table}}
\renewcommand{\thefigure}{S\arabic{figure}}
\renewcommand{\theHequation}{S\arabic{equation}}
\renewcommand{\theHtable}{S\arabic{table}}
\renewcommand{\theHfigure}{S\arabic{figure}}
\begin{widetext}
\phantomsection\label{S-sm}
\begin{center}
{\large\bfseries Supplemental Material for\\[2pt]
``Probing the CKM Origin of CP Violation with Paramagnetic EDMs \\ and Rare Kaon Decays''}\\[6pt]
Giancarlo D'Ambrosio and Luca Di Luzio
\end{center}
\vspace{4pt}
This Supplemental Material gives the derivations behind the Letter:
the matching of kaon exchange onto the paramagnetic electric dipole moment (EDM)
coefficient $\Cs$, the phases that violate combined charge conjugation and parity (CP),
and the weak nucleon couplings
(Sec.~\ref{S-app:matching}); the two-photon electron vertex and the other
electromagnetic (EM) terms (Sec.~\ref{S-app:loop}); the kaon
Cabibbo--Kobayashi--Maskawa (CKM) plane
and its uncertainty treatment (Sec.~\ref{S-app:kaonplane}); and the
new-physics matching and constraints (Sec.~\ref{S-app:NP}).
\begingroup
\let\SMcontentsline\contentsline
\def\contentsline#1#2#3#4{}
\def\SMtocstart{\let\contentsline\SMcontentsline}
\tableofcontents
\endgroup
\addtocontents{toc}{\protect\SMtocstart}
\end{widetext}

\Needspace{10\baselineskip}
\section{Kaon exchange, CP phases, and weak nucleon couplings}
\label{S-app:matching}

This section fixes the kaon field conventions, derives the
invariant CP-odd product, and determines the weak nucleon
couplings. Their real parts are fitted to hyperon decays and
their imaginary parts estimated from the $\Delta S=1$ gluonic
penguin operator $Q_6$. Sec.~\ref{S-app:loop} combines them
with the two-photon electron vertex.
\Needspace{10\baselineskip}
\subsection{Fields and matching onto the electron--nucleon interaction}

In elastic electron scattering, $t=q^2=(p'-p)^2$, where
$p,p'$ are the incoming and outgoing electron momenta, with
$p^2=p'^2=m_e^2$. For spacelike transfer we write
$Q=\sqrt{-t}$. The neutral-kaon fields are chosen as
\begin{equation}
 K_S=\frac{i(K^0-\bar K^0)}{\sqrt2},\qquad
 K_L=\frac{K^0+\bar K^0}{\sqrt2}.
 \label{S-eq:neutral-fields}
\end{equation}
With $CP\,K^0=-\bar K^0$, these are Hermitian CP-even and
CP-odd fields, respectively. They are the CP eigenstates, often denoted $K_{1,2}$; in the names of
decay modes, $K_S$ and $K_L$ denote as usual the mass eigenstates. Indirect CP violation does not alter
the CP assignment in kaon exchange at $|t|\ll m_K^2$, as shown below Eq.~\eqref{S-eq:app-cross}.
For $N=p,n$, the interaction is
$\lag\supset\sum_{j=S,L}K_j[h_j^N\bar N N-g_j^e(t)\bar e i\gamma_5 e]$,
as in Eq.~\eqref{eq:vertices-main}.

For one real kaon field, define the source
$j_K=h_K^N\bar N N-g_K^e\bar e i\gamma_5 e$. The quadratic action is
\begin{equation}
 \lag_K=-\frac12 K(\Box+m_K^2)K+Kj_K .
\end{equation}
Solving its equation of motion and substituting gives
$\lag_{\rm eff}=\frac12j_K(\Box+m_K^2)^{-1}j_K$.
The cross term is therefore
\begin{equation}
 \lag_{\rm eff}\supset
 -\frac{h_K^Ng_K^e(t)}{m_K^2-t}(\bar N N)(\bar e i\gamma_5 e),
 \label{S-eq:app-cross}
\end{equation}
which yields Eq.~\eqref{eq:KLmain}.

With both fields, $K=(K_S,K_L)$ and $j_K=(j_S,j_L)$, the quadratic action keeps this form with
$m_K^2$ replaced by a squared-mass matrix $\mathcal M^2$, symmetric because the fields are Hermitian,
and $\lag_{\rm eff}=\tfrac12 j_K^T(\mathcal M^2-t)^{-1}j_K$. Indirect CP violation is a small
off-diagonal entry $\delta$ of $\mathcal M^2$, whose eigenvalues $m_S^2$ and $m_L^2$ include the widths.
The mass eigenstates then contain admixtures of relative size $\delta/(m_L^2-m_S^2)$,
of order $\varepsilon_K$. To first order in $\delta$,
\begin{equation*}
 \begin{aligned}
 \big[(\mathcal M^2-t)^{-1}\big]_{SL}&{}=-\frac{\delta}{(m_S^2-t)(m_L^2-t)}\\
 &{}=-\frac{\delta}{m_L^2-m_S^2}\left(\frac{1}{m_S^2-t}-\frac{1}{m_L^2-t}\right).
 \end{aligned}
\end{equation*}
The admixtures therefore enter the two poles with opposite residues.
At $|t|\ll m_K^2$ the mixed term is suppressed relative to the diagonal ones by
$|m_L^2-m_S^2|/m_K^2\simeq2|\Delta m_K-\tfrac{i}{2}\Delta\Gamma_K|/m_K\simeq2\times10^{-14}$, so the CP
assignment of Eq.~\eqref{S-eq:neutral-fields} holds for the molecule. In on-shell decays,
$t=m_S^2$ or $m_L^2$, one pole is resonant and $\varepsilon_K$ contributes in full.

\Needspace{5\baselineskip}
\subsection{Electroweak and chiral vertex conventions}

In the operator expressions below we use the conventional shorthand
$\lambda_i=V_{is}^*V_{id}$ for $i=u,c,t$.
We write the EW electron operator in the form of Ref.~\cite{Ema:2022yra},
\begin{align}
 \lag_{\rm EW}&=\mathcal P_{\rm EW}
  (\bar e\gamma_\mu\gamma_5e)
  \bigl[\bar s\gamma^\mu(1-\gamma_5)d\bigr]+\mathrm{h.c.},
 \label{S-eq:PEWop}\\
 \mathcal P_{\rm EW}
 &=-\frac{G_F}{\sqrt2}\frac{\alpha_{\rm em}(m_Z)}{2\pi s_W^2}
       \lambda_tY(x_t) .
 \label{S-eq:PEW}
\end{align}
The Inami--Lim function is
\begin{equation}
 Y(x_t)=\frac{x_t}{8}
       \left[\frac{x_t-4}{x_t-1}+
                      \frac{3x_t\ln x_t}{(x_t-1)^2}\right].
\end{equation}
Here $x_t=m_t^2/m_W^2$ and $s_W=\sin\theta_W$;
$\lambda_i$ always denotes a CKM product, while the Cabibbo
parameter is written without a subscript as $\lambda$.
Throughout, we use the QCD-corrected value $Y(x_t)=0.95$ of Ref.~\cite{Dery:2025pcx}, obtained with the
running mass $\overline m_t(\overline m_t)=163\,\gev$; the leading-order (LO) expression above gives $0.94$
for this mass.
The function $I(x_t)$ in Ref.~\cite{Ema:2022yra} equals $2Y(x_t)$;
its prefactor
$\alpha_{\rm em}(m_Z)/(4\pi s_W^2)$ is thus equivalent to Eq.~\eqref{S-eq:PEW} up to the sign. The
minus sign gives the standard sign of the $Z$-penguin and box contribution~\cite{Buchalla:1998ba},
used for all semileptonic operators in Sec.~\ref{S-app:NP}; Ref.~\cite{Ema:2022yra} has the opposite
overall sign. This flips the sign of $g_S^{e,\rm EW}$ and of $C_{S,\rm EW}$ relative to that reference.
As the overall sign of the CP-conserving weak nucleon couplings is not fixed by hyperon
data~\cite{Ema:2022yra}, we compare with
Ref.~\cite{Ema:2022yra} through $|C_{S,\rm EW}|$.

We use $F_\pi=92.07\,\mev$ throughout the LO octet
calculation. The meson decay constant enters through
\begin{equation}
 \begin{aligned}
 \langle0|\bar s\gamma^\mu\gamma_5d|K^0(q)\rangle&=if_Kq^\mu,
 \end{aligned}
 \label{S-eq:axial-kaon-matching}
\end{equation}
with $f_K=\sqrt2F_\pi$ at leading chiral order.
This is the kaon-current normalization in the SU(3) approximation;
the physical value $f_K\simeq156\,\mev$ includes symmetry-breaking
corrections. In our neutral-kaon phase convention the current gives
the derivative interaction
\begin{equation}
 \begin{aligned}
 \lag_{Kee}^{\rm EW}={}&\sqrt2 f_K\left(
  \partial_\mu K_S\Ima\mathcal P_{\rm EW}
 +\partial_\mu K_L\Rea\mathcal P_{\rm EW}\right)\\
 &\hspace{3em}\times\bar e\gamma^\mu\gamma_5e .
 \end{aligned}
 \label{S-eq:derivative-kaon-electron}
\end{equation}
Integrating by parts and using the mass term in the axial Ward
identity, $\partial_\mu(\bar e\gamma^\mu\gamma_5e)\supset
2m_e\bar e i\gamma_5 e$, gives the electron bilinear with coefficient
$-2\sqrt2 f_Km_e$.
We neglect the anomalous photon term in the electron axial Ward
identity, which contributes at higher EM order in this electroweak (EW) matching.
The LO interaction is therefore
\begin{equation}
 \lag_{Kee}^{\rm EW}
 =-4F_\pi m_e\bar e i\gamma_5 e
   \left(K_S\Ima\mathcal P_{\rm EW}
                    +K_L\Rea\mathcal P_{\rm EW}\right).
 \label{S-eq:CPvertices}
\end{equation}
These formulas specify the neutral-kaon phase convention.
The bilinear $\bar e i\gamma_5 e$ is CP odd. Its product with the CP-odd $K_L$
is CP even and consequently multiplies the real weak coefficient.
Its product with the CP-even $K_S$ is CP odd and multiplies the
imaginary coefficient.

The chiral fields are fixed by the octet matrices
\begin{equation}
 M=\begin{pmatrix}
 \pi^0/\sqrt2+\eta_8/\sqrt6&\pi^+&K^+\\
 \pi^-&-\pi^0/\sqrt2+\eta_8/\sqrt6&K^0\\
 K^-&\bar K^0&-2\eta_8/\sqrt6
 \end{pmatrix},
 \label{S-eq:meson-matrix}
\end{equation}
\begin{equation}
 B=\begin{pmatrix}
 \Sigma^0/\sqrt2+\Lambda/\sqrt6&\Sigma^+&p\\
 \Sigma^-&-\Sigma^0/\sqrt2+\Lambda/\sqrt6&n\\
 \Xi^-&\Xi^0&-2\Lambda/\sqrt6
 \end{pmatrix}.
 \label{S-eq:baryon-matrix}
\end{equation}
We use
$\xi=\exp[iM/(\sqrt2F_\pi)]$, so that the meson kinetic term
is canonically normalized.

The weak and mass terms of the baryon octet Lagrangian are
\begin{equation}
 \begin{aligned}
 \lag_{\rm wB}={}&h_D\,\mathrm{Tr}\!\left(\bar B\{\xi^\dagger h\xi,B\}\right)
  \\
 &+h_F\,\mathrm{Tr}\!\left(\bar B[\xi^\dagger h\xi,B]\right)
  +\mathrm{h.c.},\\
 \lag_{\rm mass}={}&b_D\,\mathrm{Tr}\!\left(\bar B\{\mathcal M_+,B\}\right)\\
 &{}+b_F\,\mathrm{Tr}\!\left(\bar B[\mathcal M_+,B]\right)
  +b_0\,\mathrm{Tr}(\bar BB)\,\mathrm{Tr}\,\mathcal M_+,
 \end{aligned}
 \label{S-eq:baryonL}
\end{equation}
where the weak spurion $h_{ij}=\delta_{i2}\delta_{j3}$
selects the $s\to d$ transition, $h_{D,F}$ are the usual $D$- and $F$-type weak couplings,
and $\mathcal M_+=\xi^\dagger M_q\xi^\dagger+\xi M_q^\dagger\xi$ with
$M_q=\mathrm{diag}(m_u,m_d,m_s)$. At first order in $M$,
$\xi^\dagger h\xi=h+i[h,M]/(\sqrt2F_\pi)+\cdots$.
The neutral term is
$i(c_N^*K^0-c_N\bar K^0)\bar NN/(\sqrt2F_\pi)$.
Using Eq.~\eqref{S-eq:neutral-fields} therefore gives
\begin{equation}
 \begin{aligned}
 &c_p=h_D-h_F,\quad c_n=-2h_F,\\
 &h_S^N=\frac{\Rea c_N}{F_\pi},\quad
 h_L^N=\frac{\Ima c_N}{F_\pi}.
 \end{aligned}
 \label{S-eq:baryonc}
\end{equation}
The constants $h_D,h_F,c_N$ have mass dimension one.
Equivalently, before substituting $f_K=\sqrt2F_\pi$, the scalar
vertex is $h_S^N=\sqrt2\Rea c_N/f_K$, while the EW
electron vertex is $g_S^{e,\rm EW}=2\sqrt2 f_Km_e\Ima\mathcal P_{\rm EW}$.
Their product is
\begin{equation}
 h_S^Ng_S^{e,\rm EW}
 =4m_e\Rea c_N\Ima\mathcal P_{\rm EW}.
 \label{S-eq:decay-constant-cancellation}
\end{equation}
The common decay constant cancels, as emphasized in
Ref.~\cite{Ema:2022yra}.

\Needspace{5\baselineskip}
\subsection{Rephasing invariance and both placements of CP violation}

For arbitrary quark phase conventions, the invariant CKM factor is
$J=\Ima(\lambda_t\lambda_u^*)$.
In the main-text convention, $\lambda_u$ is real and
$J=\lambda_u \Ima(V_{ts}^*V_{td})$. In the $\lambda_u$-real convention, taking $h_D,h_F$ to be real
sets the scalar $K_L$--nucleon coupling to zero.
To obtain the CP-odd coupling required for two-photon $K_L$
exchange, we will estimate their CKM-induced imaginary parts
by matching $Q_6$ onto the same weak baryon Lagrangian.

Using CKM unitarity, $\lambda_u+\lambda_c+\lambda_t=0$,
we eliminate the charm CKM factor and write the electron
source and baryon coefficient as
\begin{equation}
 E=\lambda_uD_u+\lambda_tD_t,\qquad
 c_N=\lambda_u^*c_{Nu}+\lambda_t^*c_{Nt}.
 \label{S-eq:sources}
\end{equation}
Here $D_{u,t}$ and $c_{Nu},c_{Nt}$ describe the electron
and baryon vertices with the CKM factors removed; the charm
contribution is absorbed into these reduced coefficients.
For the soft matching at $t=0$, these coefficients are real,
and the weak phases are carried explicitly by the CKM factors.
The electron source is normalized by
$g_S^e=4F_\pi m_e\Ima E$ and $g_L^e=4F_\pi m_e\Rea E$
at that point, so $E=\mathcal P_{\rm EW}$ for the EW
vertex. Thus $D_{u,t}$ have dimension $-2$, whereas
$c_{Nu},c_{Nt}$ have dimension one.
Using Eqs.~\eqref{S-eq:CPvertices} and \eqref{S-eq:baryonc}, the sum of
the $K_S$ and $K_L$ pole numerators in Eq.~\eqref{eq:KLmain}, which share the
denominator $m_K^2$ at $t\simeq0$, is
\begin{align}
 h_S^Ng_S^e+h_L^Ng_L^e
 &=4m_e\Ima(Ec_N),\nonumber\\
 \Ima(Ec_N)&=J(D_tc_{Nu}-D_uc_{Nt}).
 \label{S-eq:invariant}
\end{align}
This identity displays both possible placements of CP violation and is
invariant under rephasing the strange and down quarks.

For the top EW coefficient, $D_u=0$.
The term proportional to $|\lambda_t|^2D_tc_{Nt}$ has zero
imaginary part and therefore does not contribute to the CP-odd sum.
Hence
\begin{equation}
 C_{S,\rm EW,LO}^{N}
 =\frac{2\alpha_{\rm em}(m_Z)m_e}{\pi s_W^2m_K^2}
       Jc_{Nu}Y(x_t) .
 \label{S-eq:CSinvariant}
\end{equation}
The vanishing $|\lambda_t|^2D_tc_{Nt}$ term involves both poles. With real $\lambda_u$, the
tree-level EW $K_L$ pole, $h_L^Ng_L^{e,\rm EW}=-4m_e\Rea\lambda_t\Ima\lambda_tD_tc_{Nt}$, cancels the part
of the $K_S$ pole that comes from $\Rea\lambda_tc_{Nt}$ in $\Rea c_N$. Tree-level $K_L$ exchange is
therefore contained in Eq.~\eqref{S-eq:CSinvariant} and is not a separate correction.
We define the dimensionless hadronic normalization factor
$\mathcal K_N$ through
$C_{S,\rm EW}^{N}=\mathcal K_N\Ima\lambda_tY(x_t)$.
At the accuracy of the leading real hyperon fit,
$\Rea c_N\simeq\lambda_uc_{Nu}$, this gives
\begin{equation}
 \mathcal K_N^{\rm LO}
 =\frac{2\alpha_{\rm em}(m_Z)m_e}{\pi s_W^2m_K^2}
          \Rea c_N,
 \label{S-eq:Hcoefficient}
\end{equation}
with $\mathcal K_A=[Z\mathcal K_p+(A-Z)\mathcal K_n]/A$.

For the EM correction, both weak insertions can
instead be matched from the nonleptonic operator basis. Write
\begin{equation}
 \ham_{\Delta S=1}=\frac{G_F}{\sqrt2}
 \sum_i(\lambda_u^*z_i-\lambda_t^*y_i)Q_i+\mathrm{h.c.},
 \label{S-eq:weak-operator-basis}
\end{equation}
where $Q_i$ are renormalized four-quark operators and $z_i,y_i$
are their real, dimensionless Wilson coefficients.
As in Ref.~\cite{Buchalla:1995vs}, $z_i$ multiplies
$\lambda_u$ and is dominated by the current--current operators $Q_{1,2}$, while $y_i$ multiplies
$\lambda_t$ and is nonzero only for the penguin operators; $Q_6$ dominates the CP-violating
nucleon couplings below. The electron source uses the
conjugate weak coefficients. In the notation of
Eq.~\eqref{S-eq:sources}, let
\begin{equation}
 \begin{aligned}
 E&=\sum_i(\lambda_uz_i-\lambda_ty_i)E_i,\\
 c_N&=\sum_i(\lambda_u^*z_i-\lambda_t^*y_i)B_i^N.
 \end{aligned}
 \label{S-eq:reduced-weak-sources}
\end{equation}
The real reduced quantities $E_i$ and $B_i^N$ include the weak
prefactors and the matching onto these source conventions;
their mass dimensions are $-2$ and $1$, respectively. They are
evaluated at the same dispersive matching point, after weak
phases have been extracted.

The complete CP-odd product is then
\begin{equation}
 \begin{aligned}
 &\Ima\!\left[
   \sum_i(\lambda_uz_i-\lambda_ty_i)E_i
   \sum_j(\lambda_u^*z_j-\lambda_t^*y_j)B_j^N
       \right]\\
 &\hspace{1em}=J\sum_{ij}(z_iy_j-y_iz_j)E_iB_j^N.
 \end{aligned}
 \label{S-eq:fulloperators}
\end{equation}
The right-hand side is antisymmetric under $i\leftrightarrow j$. It vanishes if a single
operator generates both vertices or, more generally, if $E_i\propto B_i^N$: the two vertices then
carry the same weak phase, and only their relative phase is physical. In this aligned case the
$K_S$ and $K_L$ terms of Eq.~\eqref{S-eq:invariant} are separately nonzero but cancel, so keeping
only one pole would produce a spurious CP-violating $\Cs$.

\Needspace{5\baselineskip}
\subsection{Hyperon fit, penguin matching, and nucleon couplings}

The CP-conserving couplings are taken from the hyperon fit of Ref.~\cite{Ema:2022yra},
\begin{equation}
 \begin{aligned}
 \Rea h_D&=-0.56\sqrt2G_FF_\pi m_{\pi^+}^2,\\
 \Rea h_F&=1.42\sqrt2G_FF_\pi m_{\pi^+}^2,
 \end{aligned}
 \label{S-eq:realab}
\end{equation}
where $m_{\pi^+}$ only sets the normalization of the fit. The fit follows the leading-order SU(3)
analysis of the $S$-wave hyperon amplitudes~\cite{Bijnens:1985kj}; chiral loops leave the fitted
couplings close to their tree-level values~\cite{Jenkins:1991bt}. These real couplings contain
CP-even current--current contributions, whereas in the convention with real $\lambda_u$ the
imaginary parts come only from the $\lambda_ty_i$ terms of Eq.~\eqref{S-eq:weak-operator-basis},
i.e.\ from penguin operators. Real and imaginary parts therefore weight different operators and,
by Eq.~\eqref{S-eq:fulloperators}, cannot be described by a single weak coefficient.

Among the penguins, $Q_6$ of Eq.~\eqref{eq:Q6operator} dominates, as it does for
$\Ima A_0$ in $K\to\pi\pi$~\cite{RBC:2020kdj}. It has the largest coefficient among the quantum chromodynamics (QCD) penguins,
$y_6=-0.113$ (LO at $\mu=1\,\gev$, as in Ref.~\cite{He:2025wxn}), and its $(V-A)(V+A)$
structure becomes a product of scalar densities after Fierzing, whose matrix elements are
chirally enhanced by $B_0\simeq2\,\gev$ (see Eq.~\eqref{S-eq:Q6fac} below).

Like the operators that generate $\lag_{\rm wB}$, $Q_6$ transforms as $(8_L,1_R)$ under
$SU(3)_L\times SU(3)_R$. Its leading baryon realization therefore has the same form,
\begin{equation}
 \begin{aligned}
 Q_6\to{}&-d_6\,\mathrm{Tr}\!\left(\bar B\{\xi^\dagger h\xi,B\}\right) -f_6\,\mathrm{Tr}\!\left(\bar B[\xi^\dagger h\xi,B]\right),
 \end{aligned}
 \label{S-eq:Q6chiral}
\end{equation}
which defines the reduced matrix elements $d_6$ and $f_6$, of dimension three. With
$\lag\supset(G_F/\sqrt2)\lambda_t^*y_6Q_6$ and $\Ima\lambda_t^*=-\Ima(V_{ts}^*V_{td})$,
comparison with Eq.~\eqref{S-eq:baryonL} gives
$\Ima(h_D,h_F)=(G_F/\sqrt2)\,\Ima(V_{ts}^*V_{td})\,y_6\,(d_6,f_6)$. The same $d_6$ and $f_6$ fix
both the hyperon and the $K^0\bar NN$ vertices.

We estimate $d_6$ and $f_6$ from baryon-to-baryon matrix elements, where no meson is
involved. The Fierz form $Q_6=-8\sum_q(\bar d_Lq_R)(\bar q_Rs_L)$ is evaluated in
factorization, by inserting the vacuum between the two densities. With
$\langle\bar qq\rangle=-F_\pi^2B_0$ and $B_0=m_K^2/[m_s+(m_u+m_d)/2]$, this gives
$Q_6\to4F_\pi^2B_0\,\bar ds$. Since $\bar ds=\bar d_Ls_R+\bar d_Rs_L$ couples to both
$(M_q)_{ds}$ and $(M_q^\dagger)_{ds}$, the mass terms of Eq.~\eqref{S-eq:baryonL} give, at
zeroth order in the meson fields,
$\bar ds\to-2[b_D\mathrm{Tr}(\bar B\{h,B\})+b_F\mathrm{Tr}(\bar B[h,B])]$; the $b_0$ term does
not contribute to $s\to d$. Hence
\begin{equation}
 d_6^{\rm fact}=8F_\pi^2B_0b_D,\qquad f_6^{\rm fact}=8F_\pi^2B_0b_F.
 \label{S-eq:Q6fac}
\end{equation}
At zeroth order in the meson fields this equals the matrix element of a weak mass term
$\bar d(1\pm\gamma_5)s$, which does not contribute to physical amplitudes (Feinberg--Kabir--Weinberg
theorem)~\cite{He:2025wxn}. The factorized $Q_6$ is not such a term: its condensate factor carries the
meson fields, so it keeps the $(8_L,1_R)$ structure of Eq.~\eqref{S-eq:Q6chiral} and contributes, as in
Refs.~\cite{Tandean:2002vy,He:2025wxn}.
Adding the bag-model contributions of Ref.~\cite{Tandean:2002vy}, with reduced matrix
elements $A_B$ and $B_B$ of dimension three, gives
\begin{equation}
 d_6=d_6^{\rm fact}+3A_B-5B_B,\qquad f_6=f_6^{\rm fact}+A_B+\tfrac{11}{3}B_B.
 \label{S-eq:Q6}
\end{equation}
For baryons, factorization is a model assumption rather than the leading term in
$1/N_c$. It dominates because it is chirally enhanced: with $b_{D,F}$ fixed by the
baryon mass splittings (at lowest order $\Sigma$--$\Lambda$ and $\Xi$--$N$),
$F_\pi^2B_0b_{D,F}\propto F_\pi^2m_K^2\Delta m_B/m_s^2$, so a $10\%$ change of $m_s$
shifts it by about $20\%$.

\begin{table}[!htbp]
 \caption{Inputs for the weak nucleon matching. The quark
 masses and the LO coefficient $y_6$ are specified at
 $\mu=1\,\gev$ in the hadronic model of Ref.~\cite{He:2025wxn}.}
 \label{S-tab:inputs}
 \footnotesize
 \begin{ruledtabular}
 \begin{tabular}{ll}
 $G_F$ & $1.1664\times10^{-5}\,\gev^{-2}$\\
 $F_\pi,m_K,m_{\pi^+}$ & $(0.09207,0.4976,0.1396)\,\gev$\\
 $\Ima(V_{ts}^*V_{td})$ & $1.42\times10^{-4}$\\
 $y_6,b_D,b_F$ & $-0.113,0.226,-0.811$\\
 $A_B,B_B$ & $(0.00140,0.00064)\,\gev^3$\\
 $m_u,m_d,m_s$ & $(2.85,6.20,123)\,\mev$\\
 \end{tabular}
 \end{ruledtabular}
\end{table}

The inputs of Table~\ref{S-tab:inputs} give
\begin{align}
 \Ima h_D&=-4.07\times10^{-12}\,\gev,\nonumber\\
 \Ima h_F&=1.36\times10^{-11}\,\gev .
\end{align}
Their propagation to the nucleon vertices is displayed in
Table~\ref{S-tab:nucleons}. The coherent average is performed on
the couplings:
\begin{equation}
 \eta_{\rm Th}
 =\frac{90h_L^p+142h_L^n}{90h_S^p+142h_S^n}
 =3.18\times10^{-4}.
 \label{S-eq:etaTh}
\end{equation}
The coherent scalar coupling is
$h_S^{\rm Th}=(90h_S^p+142h_S^n)/232=-8.05\times10^{-7}$.

The ratio $\eta_{\rm Th}$ has the size expected for a CP-violating $\Delta S=1$ hadronic
coupling. Like $\xi_0=\Ima A_0/\Rea A_0$ in $K\to\pi\pi$, it is of the form
$(\Ima\lambda_t/\lambda_u)\,R$, with $\Ima\lambda_t/\lambda_u\simeq6.5\times10^{-4}$ and $R$ the
ratio of the $y_6$-weighted $Q_6$ matrix element to the CP-conserving amplitude. Here
$R\simeq0.5$, while lattice QCD gives $|\xi_0|\simeq2.1\times10^{-4}$, i.e.\
$R\simeq0.3$~\cite{RBC:2020kdj}, an order of magnitude also confirmed by the measured
$\varepsilon'/\varepsilon$.

\begin{table}[!htbp]
 \caption{Weak nucleon couplings in the common sign convention.
 All entries are dimensionless.}
 \label{S-tab:nucleons}
 \begin{ruledtabular}
 \begin{tabular}{lrr}
 & Proton & Neutron\\
 $h_S^N$ & $-6.36\times10^{-7}$ & $-9.13\times10^{-7}$\\
 $h_L^N$ & $-1.92\times10^{-10}$ & $-2.96\times10^{-10}$\\
 $h_L^N/h_S^N$ & $3.02\times10^{-4}$ & $3.25\times10^{-4}$\\
 \end{tabular}
 \end{ruledtabular}
\end{table}

\Needspace{14\baselineskip}
\section{Two-photon contribution and EM corrections}
\label{S-app:loop}

This section follows the photon contribution from its hadronic
source to $\Cs^N$. We specify the on-shell normalization and
soft continuation, derive the electron vertex, and combine it
with the weak nucleon couplings of Sec.~\ref{S-app:matching}.
The last two subsections add the $K_S$ pole and the CP-odd pion and eta couplings.
\Needspace{5\baselineskip}
\subsection{Photon normalization and form factor}

Normalize the odd-parity photon vertex by
\begin{align}
 \lag_{K\gamma\gamma}
 &=-\frac{c_\gamma}{4}K_LF_{\mu\nu}\widetilde F^{\mu\nu},
 \label{S-eq:photon-interaction}\\
 \Gamma(K_L\to\gamma\gamma)
 &=\frac{|c_\gamma|^2m_K^3}{64\pi}.
 \label{S-eq:cgamma}
\end{align}
With virtual photons we replace $c_\gamma$ by
$F_{K\gamma\gamma}(t;k_1^2,k_2^2)$. We estimate it using
\begin{equation}
 F_{K\gamma\gamma}(t;k_1^2,k_2^2)
 =c_\gamma r_K(t)f(k_1^2,k_2^2),\qquad f(0,0)=1,
 \label{S-eq:sourcefactor}
\end{equation}
where $r_K(m_K^2)=1$ by definition and $r_K(0)=1$ for the numerical
estimate. The value $c_\gamma=+3.39\times10^{-9}\,\gev^{-1}$, from the PDG values of $\BR(K_L\to\gamma\gamma)$ and $\tau_L$~\cite{ParticleDataGroup:2024cfk}, is our nominal on-shell magnitude, with a chosen positive source sign.
The diphoton rate alone does not determine its sign relative to the
baryon coupling.

The vector-meson-dominance
(VMD) form factor of Ref.~\cite{DAmbrosio:1997eof} is
\begin{align}
 f(k_1^2,k_2^2)
 &=1+a_V\left(\frac{k_1^2}{k_1^2-M_V^2}
                  +\frac{k_2^2}{k_2^2-M_V^2}\right)\nonumber\\
 &\quad+b_V\frac{k_1^2k_2^2}
                    {(k_1^2-M_V^2)(k_2^2-M_V^2)} .
 \label{S-eq:DIP}
\end{align}
We impose $1+2a_V+b_V=0$
to ensure the QCD short-distance falloff of the form
factor~\cite{Isidori:2003ts}. This condition removes a constant
term at large, equal Euclidean photon virtualities.
For Euclidean $u>0$, with $x=u/M_V^2$, this gives
\begin{equation}
 f(-u,-u)=\frac{1+2(1+a_V)x}{(1+x)^2}.
 \label{S-eq:diagonal}
\end{equation}
The nominal inputs are $M_V=0.775\,\gev$ and $a_V=-1.611$, the combination of $K_L\to\gamma\ell^+\ell^-$ and $K_L\to e^+e^-\mu^+\mu^-$ data in Ref.~\cite{Isidori:2003ts}.
The variable $u>0$ has dimension mass squared; the
$x=u/M_V^2$ in Eq.~\eqref{S-eq:diagonal} is a local integration
variable.
The slope is imported from the decay form-factor description.
We assume that the normalized photon form factor
$f(k_1^2,k_2^2)$ is unchanged when the kaon virtuality
is continued from $t=m_K^2$ to $t=0$.
Additional vector poles can refine this prescription~\cite{Isidori:2003ts}.

\Needspace{5\baselineskip}
\subsection{Soft continuation of the photon source}

The diphoton rate fixes $|F_{K\gamma\gamma}|$ only at $(t;k_1^2,k_2^2)=(m_K^2;0,0)$. The
choice $r_K(0)=1$ in Eq.~\eqref{S-eq:sourcefactor} continues this magnitude to the soft point
$t\simeq0$ without suppression. If the vertex came only from derivative weak mixing of $K_L$ with
$\pi^0$, $\eta$ and $\eta'$, followed by their two-photon decays, it would be proportional to $t$
and would vanish at $t=0$. Weak mass-spurion operators at higher chiral order, however, give
$K_LF\widetilde F$ couplings proportional to quark masses, which stay finite as
$t\to0$~\cite{Gerard:2005yk}. Taking $r_K(0)=1$ is therefore conservative with respect to
derivative mixing.

The value at $t\simeq0$ is not fixed by on-shell data alone. A field redefinition that trades
derivative mixing for a mass term moves a constant between the kaon-pole term and a local
$\bar NNF_{\mu\nu}\widetilde F^{\mu\nu}$ coupling, leaving their sum unchanged, so only the sum is
physical. Taking $r_K(0)=1$ amounts to assuming that this sum equals the kaon-pole term with the
on-shell coupling $c_\gamma$.

\Needspace{5\baselineskip}
\subsection{Projection onto the pseudoscalar electron vertex}

Take $p'=p+q$, with $p^2=p'^2=m_e^2$ and $t=q^2$, so that
$2p\cdot q=-t$. The photons carry momenta $k$ and $q-k$
from the kaon vertex to the electron line. We use the metric
$(+,-,-,-)$, $\epsilon_{0123}=+1$, and
$\gamma_5=i\gamma^0\gamma^1\gamma^2\gamma^3$.
The interaction in Eq.~\eqref{S-eq:photon-interaction} gives the photon
Feynman rule
\begin{equation}
 -iF_{K\gamma\gamma}(t;k^2,(q-k)^2)
       \epsilon^{\mu\nu\rho\sigma}k_\rho q_\sigma .
 \label{S-eq:photon-Feynman-rule}
\end{equation}
Write the amputated loop amplitude as
$i\mathcal M=\bar u(p')\mathcal V(p',p)u(p)$.
In Feynman gauge, the two photon orderings give
\begin{equation}
 \begin{aligned}
 \mathcal V(p',p)
 &=\frac{e^2}{2}\int\frac{\dd^4k}{(2\pi)^4}
   \frac{F_{K\gamma\gamma}(t;k^2,(q-k)^2)}
        {[k^2+i0][(q-k)^2+i0]}\\
 &\quad\times\epsilon^{\mu\nu\rho\sigma}k_\rho q_\sigma
  \biggl[
 \gamma_\nu\frac{\slashed p+\slashed k+m_e}
 {(p+k)^2-m_e^2+i0}\gamma_\mu\\
 &\hspace{4em}+\gamma_\mu
 \frac{\slashed p'-\slashed k+m_e}
 {(p'-k)^2-m_e^2+i0}\gamma_\nu\biggr],
 \end{aligned}
 \label{S-eq:electron-Feynman-amplitude}
\end{equation}
where $e^2=4\pi\alpha(0)$. The factor $1/2$ compensates
for the explicit sum over the two identical-photon orderings.
The change of variable $k\to q-k$, using the symmetry of
$F_{K\gamma\gamma}$, makes the two contributions equal.
Thus either ordering can be used with the prefactor $e^2$.

For $\lag\supset-g_L^{e,\gamma\gamma}K_L\bar e i\gamma_5e$,
the on-shell Feynman vertex is $g_L^{e,\gamma\gamma}(t)\gamma_5$.
Its coefficient is extracted at nonzero $t$ by the projector
\begin{equation}
 g_L^{e,\gamma\gamma}(t)=
 \frac{\operatorname{Tr}\!\left[
  (\slashed p'+m_e)\mathcal V(p',p)
  (\slashed p+m_e)\gamma_5\right]}{2t}.
 \label{S-eq:pseudoscalar-projector}
\end{equation}
The denominator follows from the on-shell identity
\[
 \operatorname{Tr}\!\left[
 (\slashed p'+m_e)\gamma_5(\slashed p+m_e)\gamma_5\right]=2t.
\]
For the first ordering, the contracted numerator trace is
\begin{equation}
 \begin{aligned}
 &\epsilon^{\mu\nu\rho\sigma}k_\rho q_\sigma\,
 \operatorname{Tr}\!\bigl[
 (\slashed p'+m_e)\gamma_\nu
 (\slashed p+\slashed k+m_e)\\
 &\hspace{7em}\times\gamma_\mu(\slashed p+m_e)\gamma_5\bigr]\\
 &=8im_e\left[t k^2-(q\cdot k)^2\right].
 \end{aligned}
 \label{S-eq:projected-Dirac-trace}
\end{equation}
This exposes the electron-mass factor required by the chirality
flip. Using the factorized photon vertex in
Eq.~\eqref{S-eq:sourcefactor}, define the dimensionless reduced
form factor $\mathcal A(t)$ by
\begin{equation}
 g_L^{e,\gamma\gamma}(t)
   =\frac{\alpha(0)m_e}{2\pi}c_\gamma r_K(t)\mathcal A(t).
 \label{S-eq:gA}
\end{equation}
The projection gives the standard scalar loop integral
\cite{DAmbrosio:1997eof,Dorokhov:2007bd}, written here in
scattering kinematics:
\begin{equation}
 \begin{aligned}
 \mathcal A(t)
 &=\frac{2i}{\pi^2t}\int\dd^4k\,
 \frac{t k^2-(q\cdot k)^2}{[k^2+i0][(q-k)^2+i0]}\\
 &\qquad\times\frac{f(k^2,(q-k)^2)}
 {(p+k)^2-m_e^2+i0}.
 \end{aligned}
 \label{S-eq:projected-scalar-loop}
\end{equation}
The electron denominator follows directly from the momentum
$p+k$ in the first ordering of
Eq.~\eqref{S-eq:electron-Feynman-amplitude}.

We now take $t\to0^-$ after the projection. In the electron
rest frame, $p=(m_e,\boldsymbol0)$, the on-shell condition gives
$q^0=O(|\boldsymbol q|^2/m_e)$. Choosing $\boldsymbol q$
along the third axis yields
\begin{equation}
 \lim_{t\to0^-}\left[k^2-\frac{(q\cdot k)^2}{t}\right]
 =(k^0)^2-k_x^2-k_y^2,
 \label{S-eq:soft-projected-numerator}
\end{equation}
where $k_x,k_y$ are transverse to $\boldsymbol q$.
After Wick rotation, $k^0=ik_4$, put
$u=k_E^2$ and $z=k_4/\sqrt u=\cos\theta$.
After averaging the numerator over spatial directions, the
numerator and electron denominator become
\begin{equation}
 \begin{aligned}
 k^2-\frac{(q\cdot k)^2}{t}
 &\longrightarrow-\frac{u}{3}(2+z^2),\\
 (p+k)^2-m_e^2&\longrightarrow-u+2im_e\sqrt u\,z.
 \end{aligned}
 \label{S-eq:Wick-loop-kernel}
\end{equation}
Pairing $z$ with $-z$ removes the odd imaginary part of the
propagator. With the measure
$\dd^4k_E=(u/2)\dd u\,4\pi\sin^2\theta\,\dd\theta$,
the remaining angular integral, with $s(u)=\sqrt{1+4m_e^2/u}$, is
\begin{align}
 W(u)&=\frac{2}{\pi}\int_0^\pi\dd\theta\,
 \frac{\sin^2\theta\,[2+\cos^2\theta]}
                {1+(4m_e^2/u)\cos^2\theta}\nonumber\\
 &=\frac{4}{1+s(u)}+\frac{1}{[1+s(u)]^2}.
 \label{S-eq:angular}
\end{align}
The photons have opposite momenta in this limit, so their form
factor is $f(-u,-u)$. The radial integral is therefore
\begin{equation}
 \mathcal A(0)=-\frac23\int_0^\infty\frac{\dd u}{u}
                         f(-u,-u)W(u).
 \label{S-eq:radial}
\end{equation}
This retains the electron mass in the soft integration region.
As $u\to0$, $W(u)\sim2\sqrt u/m_e$, so the integral is infrared
convergent. For $m_e^2\ll u\ll M_V^2$, $W\to9/4$ and $f\simeq1$,
giving the large logarithm $3\ln(m_e/M_V)$. At large $u$, the form
factor makes the integral ultraviolet convergent.

The leading small-mass result can alternatively be written as the
subtracted inverse moment~\cite{Dorokhov:2007bd}
\begin{align}
 \mathcal A(0)&=3\ln\frac{m_e}{\mu}-\frac54
                    -\frac32 I_f(\mu)+\cdots,\nonumber\\
 I_f(\mu)&=\int_0^{\mu^2}\frac{\dd u}{u}[f(-u,-u)-1]
             +\int_{\mu^2}^\infty\frac{\dd u}{u}f(-u,-u).
 \label{S-eq:inverse}
\end{align}
For Eq.~\eqref{S-eq:diagonal},
\begin{align}
 I_f(\mu)&=1+2a_V+\ln\frac{M_V^2}{\mu^2},\nonumber\\
 \mathcal A(0)&=3\ln\frac{m_e}{M_V}-\frac{11}{4}-3a_V
 +O\!\left(\frac{m_e^2}{M_V^2}\ln^2\frac{m_e}{M_V}\right).
 \label{S-eq:Aanalytic}
\end{align}
The subtraction scale cancels explicitly.
For the benchmark VMD parameters,
$\mathcal A(0)\simeq-19.9$ and
$g_L^{e,\gamma\gamma}(0)\simeq-4.00\times10^{-14}$,
as used in Eq.~\eqref{eq:mainnumbers}.

\Needspace{5\baselineskip}

\subsection{Finite momentum transfer}

In the molecule the kaon carries spacelike momentum $t=-Q^2$, with $Q$ of nuclear size. For
$Q\ll M_V$ the loop function of Refs.~\cite{GomezDumm:1998gw,Isidori:2003ts}, continued to
$t=-Q^2$, becomes
\begin{align}
 \mathcal A(-Q^2)&=\mathcal A(0)+C(-Q^2)
                    +O(Q^2/M_V^2),\nonumber\\
 C(-Q^2)&=\frac{1}{\beta}
       \left[\operatorname{Li}_2(x)+\frac{\pi^2}{3}
                                  +\frac14\ln^2x\right],\nonumber\\
 \beta&=\sqrt{1+\frac{4m_e^2}{Q^2}},\qquad
 x=\frac{\beta-1}{\beta+1},
 \label{S-eq:Cspacelike}
\end{align}
with $C(0)=0$ and $C(-Q^2)\simeq\pi^2Q/(4m_e)$ for $Q\ll m_e$. Since $C$ is positive and
grows with $Q$, $|\mathcal A(-Q^2)|$ falls: $|\mathcal A(-Q^2)/\mathcal A(0)|\simeq0.58$ and $0.26$
at $Q\simeq5$ and $15\,\mev$. The local value $\mathcal A(0)$ used in the Letter therefore
overestimates the loop. At spacelike $t$ the loop is real; the absorptive two-photon cut of
$K_L\to\ell^+\ell^-$ appears only at timelike $t$.

In chiral perturbation theory the same loop is written with a local lepton counterterm
$\chi^r(\mu)$, as $\mathcal A(0)=\chi^r(\mu)+3\ln(m_e/\mu)-5/2$~\cite{Savage:1992ac,Valencia:1997xe}.
Our VMD form factor corresponds to $\chi^r(M_V)=-1/4-3a_V\simeq4.6$.

\Needspace{5\baselineskip}
\subsection{Numerical contribution to the electron--nucleon coefficient}

We now combine $g_L^{e,\gamma\gamma}(0)$ from
Eq.~\eqref{S-eq:gA} with $h_L^N$ from Sec.~\ref{S-app:matching}.
The LO EW comparison uses
$m_e=0.511\,\mev$, $Y(x_t)=0.95$, $\alpha_{\rm em}(m_Z)=1/128$,
$s_W^2=0.2312$, and $\Rea\lambda_t=-3.20\times10^{-4}$,
together with Table~\ref{S-tab:inputs} and $\alpha(0)=1/137.036$.
Combining with the electron loop gives
\begin{equation}
 \left|\frac{\Delta\Cs^{\rm Th,L,\gamma\gamma}}
                  {C_{S,\rm EW,LO}^{\rm Th}}\right|
 =|\eta_{\rm Th}|\left|\frac{g_L^{e,\gamma\gamma}(0)}
                                   {g_S^{e,\rm EW}}\right|
 =(3.18\times10^{-4})(35.5).
 \label{S-eq:ratioDerive}
\end{equation}
This factorization makes the numerical suppression transparent.
The corresponding coefficients are
\begin{align}
 \Delta\Cs^{p,L,\gamma\gamma}&=-3.77\times10^{-18},\nonumber\\
 \Delta\Cs^{n,L,\gamma\gamma}&=-5.81\times10^{-18},\nonumber\\
 \Delta\Cs^{\rm Th,L,\gamma\gamma}&=-5.02\times10^{-18}.
 \label{S-eq:combined-photon-numbers}
\end{align}
The common LO reference is
$|C_{S,\rm EW,LO}^{\rm Th}|=4.44\times10^{-16}$, giving $1.13\%$.
The normalization $|C_{S,\rm EW}^{\rm Th}|\simeq
6.9\times10^{-16}$, including selected higher-order corrections
\cite{Ema:2022yra}, gives $0.73\%$ for the same photon estimate.
The estimate scales linearly with $\eta_{\rm Th}$; its relative uncertainty,
dominated by the $Q_6$ baryon matrix elements, is larger than that of the leading EW contribution.

\Needspace{5\baselineskip}

\subsection{The $K_S$ pole}

The $K_S$ pole has the topology of Fig.~\ref{fig:KL-two-photon}, with the CP violation moved from
the nucleon to the photon vertex. The CP-even $K_S$ couples to nucleons through the CP-conserving
$h_S^N$, whereas its coupling to $F_{\mu\nu}\widetilde F^{\mu\nu}$, a CP-odd two-photon state, requires CP
violation in the photon source. The photon loop of Eq.~\eqref{S-eq:gA} is the same for both poles, so the
$K_S$ electron vertex is $g_S^{e,\gamma\gamma}=\rho_\gamma\,g_L^{e,\gamma\gamma}$, with $\rho_\gamma$ the ratio
of the $K_S$ and $K_L$ couplings to $F_{\mu\nu}\widetilde F^{\mu\nu}$ at the soft point. With
$h_L^N=\eta_N h_S^N$, Eq.~\eqref{eq:KLmain} gives
\begin{align}
 \Delta\Cs^{N,\gamma\gamma,\mathrm{pole}}
 &=-\frac{\sqrt2 h_S^Ng_L^{e,\gamma\gamma}}{G_Fm_K^2}
             (\eta_N+\rho_\gamma),\nonumber\\
 \Delta\Cs^{\rm Th,\gamma\gamma,\mathrm{pole}}
 &=-5.02\times10^{-18}-1.58\times10^{-14}\rho_\gamma ,
 \label{S-eq:partnerAppendix}
\end{align}
where the first number is the $K_L$ pole of Eq.~\eqref{S-eq:combined-photon-numbers} and the
coefficient of $\rho_\gamma$ is this number divided by $\eta_{\rm Th}$.

Each pole alone depends on the kaon phase convention; only $\eta_N+\rho_\gamma$ does not: a
rephasing of the $s$ or $d$ quark multiplies $c_N$ and $E$ of Eq.~\eqref{S-eq:sources} by opposite phases
and shifts $\eta_N$ and $\rho_\gamma$ by opposite amounts. With real
$\lambda_u$, $\rho_\gamma$ is the weak phase of the $K^0\to\gamma\gamma$ amplitude that gives
$K_L\to\gamma\gamma$~\cite{DAmbrosio:1996lam}. It is not the CP violation of on-shell
$K_S\to\gamma\gamma$, which is dominated by the $\varepsilon_K$ admixture; that admixture cancels between
the poles at $t\simeq0$ (Sec.~\ref{S-app:matching}). In the notation of
Eq.~\eqref{S-eq:reduced-weak-sources}, $\eta_N=(\Ima\lambda_t/\lambda_u)\,r_N$ and
$\rho_\gamma=-(\Ima\lambda_t/\lambda_u)\,r_\gamma$, where $r_N=\sum_iy_iB_i^N/\sum_iz_iB_i^N$ and
$r_\gamma=\sum_iy_iE_i/\sum_iz_iE_i$ weigh the top-quark against the up-quark coefficients at the nucleon
and photon vertices. Hence $\eta_N+\rho_\gamma=(\Ima\lambda_t/\lambda_u)(r_N-r_\gamma)$, which is
Eq.~\eqref{S-eq:fulloperators} for these two vertices: the two poles cancel if the vertices have the same
operator content, $r_N=r_\gamma$.

The photon vertex carries a much smaller weak phase than the nucleon vertex. In
$K_L\to\gamma\gamma$ the weak interaction converts the kaon into neutral $\bar qq$ states, which reach two
photons through the anomaly with weight $e_q^2$. In large-$N_c$ chiral perturbation theory with the
$\eta'$~\cite{Gerard:2005yk}, the neutral mesons are the flavour states $\bar uu$, $\bar dd$ and $\bar ss$,
with $m_{\bar qq}^2=2B_0m_q$ at LO. The penguins $Q_4$ and $Q_6$ convert the kaon into $\bar dd$ and
$\bar ss$ with equal amplitudes. Since $m_K^2-m_{\bar dd}^2=m_{\bar ss}^2-m_K^2=B_0(m_s-m_d)$, the two
poles have opposite signs at $q^2=m_K^2$, and as $e_d=e_s$ the two cancel at LO. Only the $\bar sd\to\bar uu$ transition remains,
proportional to $x_1+x_3-x_5$ with $x_i$ the low-energy weights of $Q_i$, so that
$r_\gamma=(y_3-y_5)/(x_1+x_3-x_5)$ and
\begin{equation}
 \rho_\gamma=-\frac{\Ima\lambda_t}{\lambda_u}\,\frac{y_3-y_5}{x_1+x_3-x_5}\simeq+2.1\times10^{-5}.
 \label{S-eq:rhoGST}
\end{equation}
Here $x_i=z_i$ in factorization, with the coefficients of Eq.~\eqref{S-eq:weak-operator-basis}
at LO, $z_1=-0.743$, $y_3=0.0375$ and $y_5=0.0133$ at $\mu=1\,\gev$~\cite{Buchalla:1995vs}; the
fitted $x_1\simeq-1.3$ of Ref.~\cite{Gerard:2005yk} gives $1.2\times10^{-5}$. The nucleon vertex,
by contrast, is dominated by $Q_6$, whose scalar densities have chirally enhanced baryon matrix elements
(Sec.~\ref{S-app:matching}): $r_N\simeq0.5$, against $|r_\gamma|\simeq0.03$. Thus $\rho_\gamma$ is
$4$--$7\%$ of $\eta_{\rm Th}$, with the same sign, and the $K_S$ pole raises the $K_L$ term from
$0.73\%$ to about $0.76\%$ of the value of Ref.~\cite{Ema:2022yra}. Even if the photon vertex carried the weak phase of the isospin-zero $K\to\pi\pi$ amplitude,
$\rho_\gamma=\xi_0\simeq-2.1\times10^{-4}$~\cite{RBC:2020kdj}, the pole term would only drop to
$0.25\%$.

These estimates assume, as for $g_L^{e,\gamma\gamma}$, that the soft photon vertex has the
operator content of the on-shell one. A larger penguin share at $t\simeq0$ does not change the
size of the pole term qualitatively. If $Q_6$ entered the soft vertex with a matrix element $k$ times
that of $Q_1$, so that $r_\gamma=(y_3-y_5+ky_6)/(x_1+x_3-x_5+kz_6)$, Eq.~\eqref{S-eq:partnerAppendix}
would give between about $0.8\%$ ($k=0$) and $-1.2\%$ ($k=10$) of the value of
Ref.~\cite{Ema:2022yra}, and the two poles would cancel for $k\simeq3.4$, where the photon vertex has
the penguin share of the nucleon vertex, $r_\gamma=r_N$. A sign-flipped pole term of order $10\%$ would
require $k\gtrsim200$, a vertex generated by $Q_6$ alone, including its CP-conserving part. Such a vertex
is not expected: at LO, $Q_6$ converts the kaon into $\bar dd$ and $\bar ss$ either through derivative
terms, which vanish at $t=0$ and cancel at $t=m_K^2$, or through a weak mass term, which a rotation in
the $d$--$s$ plane removes without changing the photon couplings, since $e_d=e_s$. We therefore adopt
the on-shell operator content, as for $c_\gamma$, which gives the $0.76\%$ quoted in the Letter.

\Needspace{5\baselineskip}
\subsection{CP-odd pion and eta couplings}

We now evaluate the EM terms induced by the CP-odd
pion-- and eta--nucleon couplings
\begin{equation}
 \lag_{\rm CP}=\bar g_0\,\bar N\tau^aN\pi^a+\bar g_1\,\bar NN\pi^0+\bar g_0^\eta\,\bar NN\eta,
 \label{S-eq:gbarL}
\end{equation}
with isoscalar ($\bar g_0$, $\bar g_0^\eta$) and isovector ($\bar g_1$) couplings. They generate
$\Cs$ in two ways~\cite{Flambaum:2019ejc,Mulder:2025esr}. In meson exchange, the
$\pi^0$ or $\eta$ emitted at the CP-odd vertex couples to $\bar ei\gamma_5e$ through its anomalous
two-photon vertex and a photon loop on the electron line, as for the kaon in
Fig.~\ref{fig:KL-two-photon}(b). This is the diagram first drawn, but not evaluated, in Fig.~3b of
Ref.~\cite{Pospelov:2013sca}. In the pion loop, a charged pion emitted at the $\bar g_0$ vertex is
reabsorbed after the nucleon exchanges a photon through its magnetic moment; the resulting CP-odd
nucleon--two-photon coupling again reaches the electron through a photon loop. For a nucleus with $Z$
protons and $N$ neutrons, Ref.~\cite{Mulder:2025esr} finds
\begin{equation}
 \begin{aligned}
 \Cs={}&2.18\Big(\bar g_1+\frac{Z-N}{A}\,\bar g_0\Big)+0.0615\,\bar g_0^\eta\\
 &-\bar g_0\Big(0.391\frac ZA+0.571\frac NA\Big).
 \end{aligned}
 \label{S-eq:MulderCs}
\end{equation}
The first two terms are the meson exchange. Its $\pi^0$ part has the form of the kaon pole: with
$B^K\equiv-2\mathcal A(0)=39.8$, Eqs.~\eqref{eq:KLmain} and \eqref{eq:mainnumbers} give
$\Delta\Cs^{L,\gamma\gamma}=(\sqrt2/G_F)\,h_L^N\alpha m_ec_\gamma B^K/(4\pi m_K^2)$, and the $\pi^0$ term
follows for $h_L^N\to\bar g$, $c_\gamma\to\alpha/(\pi F_\pi)$, the anomalous $\pi^0\gamma\gamma$ coupling,
$m_K\to m_\pi$ and $B^K\to B^\pi$, so that
$2.18=(\sqrt2/G_F)\,\alpha^2m_eB^\pi/(4\pi^2F_\pi m_\pi^2)$. Here
$B^\pi=43.8$ and $B^\eta=45.8$ are the two-photon loop functions of the $\pi^0$ and $\eta$ at zero
momentum transfer, with the lepton counterterm fitted to $\pi^0\to e^+e^-$ and $\eta\to\mu^+\mu^-$.
The $\eta$ term differs by its anomalous coupling and mass,
$0.0615=2.18\,(F_\pi/\sqrt3F_\eta)(m_\pi^2/m_\eta^2)(B^\eta/B^\pi)$ for $F_\eta\simeq1.3F_\pi$. The last term
is the pion loop, whose coefficients are
$(\sqrt2/G_F)(\alpha^2/4\pi)\,m_eg_A|\mu_{n,p}|[\ln(m_\pi/m_e)+1.77]/(\mu_NF_\pi m_\pi m_N)$, with the
neutron magnetic moment $\mu_n$ for protons and the proton moment $\mu_p$ for neutrons.

In the Standard Model (SM) the couplings arise at second order in the weak interaction: a nucleon turns into a
hyperon through the CP-conserving $\Delta S=1$ hyperon--nucleon transition and back through a CP-odd
$\Delta S=1$ meson--baryon vertex, which emits the meson. The CP-odd vertex comes from $Q_5$ and $Q_6$,
evaluated in vacuum saturation. Ref.~\cite{Yamanaka:2015ncb} finds $\bar g_0=-1.1\times10^{-17}$,
$\bar g_1=-1.3\times10^{-17}$ and $\bar g_0^\eta=1.8\times10^{-17}$, with an $O(100\%)$ uncertainty
from vacuum saturation. Like the kaon terms, they are proportional to $J$, so their ratio to the EW
term does not depend on the CKM inputs, and their intermediate states are hyperons, so they do not
double count the kaon poles. For thorium, $(Z-N)/A=-0.224$ and
Eq.~\eqref{S-eq:MulderCs} gives
\begin{align}
 \Delta\Cs^{\pi^0,\eta}&=-2.19\times10^{-17},\nonumber\\
 \Delta\Cs^{\pi^\pm\ {\rm loop}}&=+5.51\times10^{-18},
 \label{S-eq:othermesons}
\end{align}
The isovector $\eta$ coupling $\bar g_1^\eta=1.6\times10^{-17}$ of Ref.~\cite{Yamanaka:2015ncb}, not
included in Eq.~\eqref{S-eq:MulderCs}, would change the $\pi^0,\eta$ term by $1\%$. The sum,
$-1.63\times10^{-17}$, is $3.7\%$ of the LO EW result and about $2.4\%$ of the value of
Ref.~\cite{Ema:2022yra}, as quoted in the Letter. These terms, which involve no kaon pole, are
thus the largest EM correction, about three
times the kaon-pole term.

The two terms have the same form but place the weak interactions differently. Per unit CP-odd
coupling the
kaon pole is weaker by $(c_\gamma/m_K^2)/[\alpha/(\pi F_\pi m_\pi^2)]\simeq1\times10^{-8}$, because its
two-photon vertex is weak and its propagator heavier. Its CP-odd coupling, however, is first order in
the weak interaction, $|h_L^{\rm Th}|=2.6\times10^{-10}$, whereas $\bar g\sim10^{-17}$ is second order.
Both corrections are therefore second order in $G_F$, proportional to $J$, and of order $\alpha^2$, so
they come out comparable: about $1\%$ and $4\%$ of the LO EW result.
\Needspace{14\baselineskip}
\section{Kaon CKM plane and experimental sensitivities}
\label{S-app:kaonplane}

This section collects the formulae and uncertainties used in
Fig.~\ref{fig:kaon-plane}. We define the kaon coordinates,
normalize the three CP-sensitive observables, and give the
charged-kaon and mixing constraints. We also inspect the precision required for
a correlated molecular EDM--kaon test.

\Needspace{5\baselineskip}
\subsection{Coordinates and CKM conversion}

For compact expressions in this section, write
\begin{equation}
 x=A^2(1-\hat\rho),\qquad y=A^2\hat\eta,
 \qquad \lambda_i=V_{is}^*V_{id}.
 \label{S-eq:plane-xy}
\end{equation}
The definition in Eq.~\eqref{eq:kaon-plane} implies
$r\equiv\lambda_t/\lambda_c=\lambda^4(x-iy)$.
In the convention $\lambda_u>0$, unitarity gives
\begin{equation}
 \lambda_c=-\frac{\lambda_u}{1+r},\qquad
 \lambda_t=-\frac{\lambda_u r}{1+r}.
 \label{S-eq:plane-exact-CKM}
\end{equation}
With $D=|1+r|^2=(1+\lambda^4x)^2+\lambda^8y^2$,
\begin{align}
 \Ima\lambda_t&=\frac{\lambda_u\lambda^4y}{D},\nonumber\\
 \Rea\lambda_t&=-\frac{\lambda_u\lambda^4
                 [x+\lambda^4(x^2+y^2)]}{D}.
 \label{S-eq:plane-ReIm}
\end{align}
The rephasing invariant entering the weak kaon--nucleon couplings is
$J=\Ima(\lambda_t\lambda_u^*)=|\lambda_c|^2\lambda^4y$.
The normalization $\lambda_u=|V_{ud}V_{us}|$ does not require a
$B$-decay input. In the numerical illustration we use
$\lambda_u=\lambda\sqrt{1-\lambda^2}$, neglecting the tiny
$|V_{ub}|^2$ correction to first-row normalization.
No value of $|V_{cb}|$ is imposed.

At leading Wolfenstein order,
\begin{equation}
 \lambda_t\simeq\lambda^5(-x+iy),\qquad
 x\simeq A^2(1-\bar\rho),\quad y\simeq A^2\bar\eta.
 \label{S-eq:plane-Wolfenstein}
\end{equation}
The hatted kaon coordinates and the conventional barred
$bd$-triangle coordinates agree only to the stated order.
For example, at the SM benchmark $(x,y)=(0.57,0.2535)$ of Fig.~\ref{fig:kaon-plane}, the exact conversion changes
$\Ima\lambda_t$ by about $-2.8\%$ relative to $\lambda^5y$.
This known normalization correction is retained in the figure.
The familiar circular and hyperbolic shapes below are their
LO limits.

\Needspace{5\baselineskip}
\subsection{Molecular EDM, muon, and neutrino determinations}

The absolute normalizations replacing the proportionalities in
Eq.~\eqref{eq:plane-CP-probes} are
\begin{align}
 C_{S,\rm EW}^A
   &=\mathcal K_A\Ima\lambda_t Y(x_t),
 \label{S-eq:masterelectron}\\
 g_{S0}^{\mu,\rm SD}
   &=\mathcal N_\mu\Ima\lambda_t Y(x_t),
 \label{S-eq:mastermuon}\\
 \BR_L\equiv\BR(K_L\to\pi^0\nu\bar\nu)
   &=\kappa_L\left[\frac{\Ima\lambda_t X(x_t)}{\lambda^5}\right]^2.
 \label{S-eq:masterneutrino}
\end{align}
The superscript SD denotes the short-distance part of the muon amplitude.
The coupling $g_{S0}^{\mu,\rm SD}$ normalizes the spin-singlet
pseudoscalar decay amplitude. For rates its modulus equals that
of $g_S^{\mu,\rm SD}(m_K^2)$ in the interaction convention of
Eq.~\eqref{eq:vertices-main}.
Here $\mathcal K_A$ is the hadronic normalization factor derived in
Eq.~\eqref{S-eq:Hcoefficient}. We identify $\Cs^A$ with $C_{S,\rm EW}^A$: the
EM terms of Sec.~\ref{S-app:loop} are at the few-percent level, well below the
weak-nucleon uncertainty. The molecular EDM energy shift is
$d_{e,a}^{\rm eq}=d_e+\zeta_a\Cs^A$.
For the CKM interpretation $d_e$ is negligible; additional EDM
sources require the joint analysis discussed in Sec.~\ref{S-app:NP}.

The muon normalization is
$\mathcal N_\mu=G_F\alpha_{\rm em}(m_Z)f_Km_\mu/(\pi s_W^2)$,
where the physical $f_K$ is defined by
$\langle0|\bar s\gamma_\mu\gamma_5d|K^0(p)\rangle=if_Kp_\mu$.
The CP-violating partial branching ratio is
\begin{equation}
 \BR_S^{(0)}=\frac{\tau_S}{\hbar}
      \frac{m_K\beta_\mu}{8\pi}|g_{S0}^{\mu,\rm SD}|^2,
 \qquad \beta_\mu=\sqrt{1-4m_\mu^2/m_K^2}.
 \label{S-eq:Swave-branching}
\end{equation}
The tagged amplitude and its mixing corrections
must be converted to this convention consistently
\cite{Dery:2021mct,Brod:2022khx}.
The neutral-neutrino normalization is
$\kappa_L=(2.231\pm0.013)\times10^{-10}(\lambda/0.225)^8$
\cite{Buras:2015qea}; it already includes the neutrino-flavour sum.

Equations~\eqref{S-eq:masterelectron}--\eqref{S-eq:Swave-branching}
combined with Eq.~\eqref{S-eq:plane-ReIm} define the three bands.
They are nearly horizontal in the exact coordinates.
Rate data determine $|y|$; the figure retains both sign branches.

\Needspace{5\baselineskip}
\subsection{Charged-kaon, mixing, and dispersive dimuon constraints}

The standard branching-ratio expression is
\cite{Buras:2015qea}
\begin{align}
 \BR_+&\equiv\BR(K^+\to\pi^+\nu\bar\nu)\nonumber\\
 &=\kappa_+(1+\Delta_{\rm EM})
 \left\{\left[\frac{\Ima\lambda_t X(x_t)}{\lambda^5}\right]^2
 \right.\nonumber\\
 &\hspace{14mm}\left.
 +\left[\frac{\Rea\lambda_c}{\lambda}P_c
       +\frac{\Rea\lambda_t}{\lambda^5}X(x_t)\right]^2\right\}.
 \label{S-eq:plane-Kplus}
\end{align}
Here $\Delta_{\rm EM}$ denotes the EM correction.
The effective charm quantity $P_c$ includes its small
long-distance correction. The tiny imaginary charm amplitude is
neglected at this accuracy. The exact coordinate conversion in
Eq.~\eqref{S-eq:plane-exact-CKM} is retained in the numerical evaluation.
The leading Wolfenstein form is
\begin{equation}
 \BR_+\simeq\kappa_+(1+\Delta_{\rm EM})X(x_t)^2
       \left[y^2+\left(x+\frac{P_c}{X(x_t)}\right)^2\right].
 \label{S-eq:plane-Kplus-circle}
\end{equation}
The charm amplitude therefore shifts the annulus centre to
$x=-P_c/X(x_t)$.
Using the input set of Ref.~\cite{Dery:2025pcx},
\begin{align}
 \kappa_+&=(5.173\pm0.025)\times10^{-11}
                         (\lambda/0.225)^8,\nonumber\\
 P_c&=[(0.3604\pm0.0087)+(0.04\pm0.02)]
                         (0.225/\lambda)^4,\nonumber\\
 X(x_t)&=1.462\pm0.017,\qquad\Delta_{\rm EM}=-0.003.
 \label{S-eq:plane-Kplus-inputs}
\end{align}
At $\lambda=0.225$ the coefficient in
Eq.~\eqref{S-eq:plane-Kplus-circle} is $1.10\times10^{-10}$ and
the charm shift is $0.274$.
The measured input is the NA62 2016--2024 combination
reported in 2026,
$\BR_+=(9.6^{+1.9}_{-1.8})\times10^{-11}$,
with a relative precision of approximately $20\%$
\cite{NA62:2026rwr}.

Kaon mixing provides a complementary constraint.
The observable $\varepsilon_K$ probes the CP-odd part of
$K^0$--$\bar K^0$ mixing. In the $u$--$t$ organization of the
short-distance amplitude~\cite{Brod:2019rzc}, define
\begin{align}
 S_{tt}&=S_0(x_t)+S_0(x_c)-2S_0(x_c,x_t),\nonumber\\
 S_{ut}&=S_0(x_c)-S_0(x_c,x_t),\nonumber\\
 T&=\eta_{tt}S_{tt},\qquad U=\eta_{ut}S_{ut},
 \label{S-eq:plane-mixing-functions}
\end{align}
where $x_i=\overline m_i^2/M_W^2$ with the mass conventions
specified below. The QCD corrections are contained in
$\eta_{tt},\eta_{ut}$; $S_{ut}$ is negative.
The all-quadrant constraint is
\begin{equation}
 |\varepsilon_K|=\frac{\kappa_\varepsilon C_\varepsilon
                            \widehat B_K}{2}
       \left|\Ima\left(\lambda_t^2T+
                        2\lambda_u\lambda_t U\right)\right|,
 \label{S-eq:plane-epsilon}
\end{equation}
with
\begin{equation}
 C_\varepsilon=
 \frac{G_F^2f_K^2m_KM_W^2}{6\sqrt2\pi^2\Delta m_K}.
 \label{S-eq:plane-Cepsilon}
\end{equation}
The bag parameter $\widehat B_K$ normalizes the $\Delta S=2$
matrix element; $\kappa_\varepsilon$ accounts for the usual
long-distance and phase corrections. The outer absolute value
must be retained for either sign of $x$ or $y$.
The loop functions, expanded to first order in $x_c$ for the
mixed contribution, are
\begin{align}
 S_0(z)&=\frac{z(4-11z+z^2)}{4(1-z)^2}
        -\frac{3z^3\ln z}{2(1-z)^3},\nonumber\\
 S_0(x_c,x_t)&=x_c\left[\ln\frac{x_t}{x_c}
       -\frac{3x_t}{4(1-x_t)}\right.\nonumber\\
       &\hspace{10mm}\left.
       -\frac{3x_t^2\ln x_t}{4(1-x_t)^2}\right]+O(x_c^2).
 \label{S-eq:plane-InamiLim}
\end{align}
Their leading coordinate dependence is
\begin{equation}
 |\varepsilon_K|\simeq
 \kappa_\varepsilon C_\varepsilon\widehat B_K\lambda^{10}T
   \left|y\left(x-\frac{U}{\lambda^4T}\right)\right|.
 \label{S-eq:plane-epsilon-hyperbola}
\end{equation}
Since $U<0$, the hyperbola is shifted in the same real-axis
sense as the charged-neutrino annulus, with a different offset.
The bands in Fig.~\ref{fig:kaon-plane} use
Eq.~\eqref{S-eq:plane-epsilon}, retaining the exact CKM conversion.
The benchmark inputs are
\begin{align}
 \kappa_\varepsilon&=0.94\pm0.02,&
 \widehat B_K&=0.7625\pm0.0097,\nonumber\\
 \eta_{tt}&=0.55(1\pm0.042),&
 \eta_{ut}&=0.404(1\pm0.013),
 \label{S-eq:plane-epsilon-inputs}
\end{align}
following Ref.~\cite{Dery:2025pcx} and its primary theory
reference~\cite{Brod:2019rzc}.
The measured magnitude is
$|\varepsilon_K|=(2.228\pm0.011)\times10^{-3}$
\cite{ParticleDataGroup:2024cfk}.

An additional real-amplitude constraint comes from dimuons.
The total $K_L$ dimuon rate also probes the real quark amplitude.
In the diphoton-normalized convention
\cite{Isidori:2003ts,Hoferichter:2023wiy},
\begin{align}
 \BR(K_L\to\mu^+\mu^-)
 &=\BR(K_L\to\gamma\gamma)\,2\beta_\mu
        \left(\frac{\alpha(0)m_\mu}{\pi m_K}\right)^2\nonumber\\
 &\quad\times\left\{(\Ima\mathcal A_{\rm LD})^2+
       [\Rea\mathcal A_{\rm LD}+\chi_{\rm short}]^2\right\},
 \label{S-eq:plane-KLmu}
\end{align}
where $\mathcal A_{\rm LD}$ is the long-distance amplitude, dominated by two-photon
exchange~\cite{GomezDumm:1998gw,Isidori:2003ts}, and the
short-distance term is
\begin{equation}
 \chi_{\rm short}=-\widetilde\kappa
       [\Rea\lambda_tY(x_t)+\Rea\lambda_cY_{\rm NL}].
 \label{S-eq:plane-chi}
\end{equation}
The dimensionless factor $\widetilde\kappa$ converts the
short-distance amplitude to this diphoton normalization;
$Y_{\rm NL}$ is the charm contribution, including its QCD
corrections. The imaginary part of $\mathcal A_{\rm LD}$ here
is absorptive and need not carry a CP-violating weak phase.
The numerical illustration in Ref.~\cite{Dery:2025pcx} uses
$\widetilde\kappa=4988$, $Y(x_t)=0.95$,
$Y_{\rm NL}=2.8\times10^{-4}$,
$\Ima\mathcal A_{\rm LD}\simeq-5.2$, and the two dispersive
solutions $\Rea\mathcal A_{\rm LD}=0.16\pm0.38$ or
$-0.66\pm0.38$.
For each solution, solving the rate for the real total amplitude
gives two square-root branches.
Tagged interference can correlate a branch with
the sign of the imaginary coordinate
\cite{Dery:2022yqc,Dery:2025pcx}.
Fig.~\ref{fig:kaon-plane} displays this constraint for both
dispersive solutions, which correspond to the two possible signs of
the $K_L\to\gamma\gamma$ amplitude relative to the short-distance
contribution, whose sign is fixed in the SM; neither is
selected here. We use
$\BR(K_L\to\mu^+\mu^-)=(6.84\pm0.11)\times10^{-9}$ and
$\BR(K_L\to\gamma\gamma)=(5.47\pm0.04)\times10^{-4}$~\cite{ParticleDataGroup:2024cfk}
and $\Ima\mathcal A_{\rm LD}=-5.21$ from two-photon unitarity, as in the
minimal-flavour-violation (MFV) analysis of Sec.~\ref{S-app:NP}, and treat $\Rea\mathcal A_{\rm LD}$
as a Gaussian nuisance parameter: each region is the $95\%$ region
of the rate $\chi^2$ minimized over $\Rea\mathcal A_{\rm LD}$, and its
width is dominated by the dispersive uncertainty. The measured rate
gives $|\Rea\mathcal A_{\rm LD}+\chi_{\rm short}|=1.12$. The two
solutions differ by $0.82$ in $\Rea\mathcal A_{\rm LD}$, which
displaces their bands by $0.31$ in $A^2(1-\hat\rho)$. At the SM
benchmark the predicted rate agrees with the measurement within
$0.1\sigma$ for $\Rea\mathcal A_{\rm LD}=-0.66$ and differs from it by
$2.0\sigma$ for $\Rea\mathcal A_{\rm LD}=0.16$; Ref.~\cite{Hoferichter:2023wiy} finds
$1.7\sigma$ for this solution.
Direct CP violation in $K\to\pi\pi$, through
$\varepsilon'/\varepsilon$, can supply further information but
requires a separate treatment of its hadronic uncertainty
\cite{DAmbrosio:2023irq}.

\Needspace{5\baselineskip}
\subsection{Numerical inputs, uncertainty treatment, and future precision}

Fig.~\ref{fig:kaon-plane} uses the theory inputs
of Ref.~\cite{Dery:2025pcx}, together with the
2026 NA62 branching ratio
and $|\varepsilon_K|$ measurements quoted above, as well as the $K_L$ dimuon and diphoton
rates. The star marks a chosen SM benchmark at $(x,y)=(0.57,0.2535)$,
which sets the central values of all three projected CP measurements.
It is not a fitted CKM determination and imposes no $B$-physics
constraint. The existing measured constraints and the hypothetical
future determinations are distinguished by separate legends and
solid versus broken boundaries in both panels. The one exception is the dotted boundary
of the $K_L\to\mu^+\mu^-$ solution with $\Rea\mathcal A_{\rm LD}=-0.66$, which is listed with the
existing constraints.

The remaining central inputs are
$\lambda=0.22501$, $\overline m_t(\overline m_t)=163\,\gev$,
$\overline m_c(\overline m_c)=1.27\,\gev$,
$M_W=80.38\,\gev$, $f_K=0.1557\,\gev$,
$m_K=0.4976\,\gev$, and
$\Delta m_K=3.484\times10^{-15}\,\gev$.
For the EW and muon normalizations we use
$Y(x_t)=0.95$, $\alpha_{\rm em}(m_Z)=1/128$,
$s_W^2=0.2312$, and $m_\mu=0.1057\,\gev$, also in the LO matching of Sec.~\ref{S-app:loop}.
For the plane,
the electron contribution is normalized to the value
$|C_{S,\rm EW}^{\rm Th}|=6.9\times10^{-16}$ of Ref.~\cite{Ema:2022yra} at
$\Ima\lambda_t=1.42\times10^{-4}$ and rescaled by the exact
$\Ima\lambda_t$ at every point.
The benchmark has $\Ima\lambda_t=1.42\times10^{-4}$, $\BR_S^{(0)}=1.89\times10^{-13}$ and
$\BR(K_L\to\pi^0\nu\bar\nu)=2.89\times10^{-11}$. The projected bands of Fig.~\ref{fig:kaon-plane}
depend only on ratios to the benchmark, so these normalizations do not affect them.

The electron projection assumes a hypothetical $20\%$ molecular EDM
measurement and retains a $30\%$ weak-nucleon normalization error,
giving $\sqrt{0.20^2+0.30^2}=36\%$ at amplitude level.
The muon illustration uses $35\%$ amplitude sensitivity
\cite{DAmbrosio:2025mxa}, while the neutrino illustration uses
$25\%$ rate precision~\cite{KOTO:2025gvq}.

The observable predictions are evaluated using the exact CKM
conversion in Eqs.~\eqref{S-eq:plane-exact-CKM} and
\eqref{S-eq:plane-ReIm}. The shaded regions in
Fig.~\ref{fig:kaon-plane} are illustrative Gaussian pull bands
with nominal $95\%$ coverage, defined by a pull of magnitude below
$1.96$. For the charged mode, the asymmetric experimental errors
are combined in quadrature with local propagation of independent
$X$, $P_c$, and $\kappa_+$ uncertainties. For
$|\varepsilon_K|$, the experimental error is combined with a
$5\%$ relative theory uncertainty. The $K_L\to\mu^+\mu^-$ regions use the profiled
rate $\chi^2$ described above, with the same $95\%$ threshold. The projected CP measurements
use central short-distance inputs and the experimental and
weak-nucleon uncertainties specified above.

\Needspace{14\baselineskip}
\section{New-physics operators, benchmarks, and kaon bounds}
\label{S-app:NP}

This section collects the operators and constraints behind the new-physics examples of the
Letter. We first discuss direct EW sources of an electron EDM, the semileptonic operators with the
apparent CKM determinations they imply, and the direct electron--kaon bound. We then give the
neutral-vector and leptoquark benchmarks, the reach on MFV operators, and light mediators and scalar
operators.
\Needspace{5\baselineskip}
\subsection{Direct EW sources of an electron EDM}

The Higgs--top estimate in the main text uses the leading
photon--Higgs Barr--Zee contribution of Ref.~\cite{Brod:2013cka},
with the electron Yukawa coupling fixed to its SM value.
The HfF$^+$ limit alone then implies
$|\widetilde\kappa_t|\lesssim4.6\times10^{-4}$, assuming no
competing EDM contributions.

A contribution closer to the CKM molecular EDM scale is provided by
the EW-multiplet example of Ref.~\cite{Banno:2024apv}.
Two new fermion doublets $\psi_A,\psi_B$ and a complex scalar
singlet $S$ interact through
\begin{equation}
 \lag\supset-\bar\psi_B(s_{\rm EW}+a_{\rm EW}\gamma_5)\psi_A S
 +\mathrm{h.c.},
 \label{S-eq:EWmultipletYukawa}
\end{equation}
where $s_{\rm EW},a_{\rm EW}$ are complex Yukawa couplings.
These couplings
generate a CP-odd operator containing three $SU(2)_L$ field
strengths at two loops, which induces $d_e$ through a further
loop. For $m_A=1\,\mathrm{TeV}$, $m_B=m_S=10\,\mathrm{TeV}$,
and $\Ima(s_{\rm EW}a_{\rm EW}^*)=0.25$, the result is
\begin{equation}
 |d_e|\simeq2.2\times10^{-34}\,\ecm,
 \label{S-eq:EWmultipletbenchmark}
\end{equation}
about 21 times the CKM equivalent EDM in ThO. This mechanism
requires no new strangeness-changing coupling.
Both benchmarks assume CP-violating phases beyond CKM.

\Needspace{5\baselineskip}
\subsection{Semileptonic normalization and gauge-invariant operators}

We write the effective neutral-current Hamiltonian as
\begin{align}
 \ham_{\rm NC}={}&
 (a_L^\ell\,\bar s\gamma_\mu P_Ld+
  a_R^\ell\,\bar s\gamma_\mu P_Rd)
                    \bar\ell\gamma^\mu\gamma_5\ell\nonumber\\
 &+(n_L^i\,\bar s\gamma_\mu P_Ld+
  n_R^i\,\bar s\gamma_\mu P_Rd)
                    \bar\nu_i\gamma^\mu P_L\nu_i
 +\mathrm{h.c.}.
 \label{S-eq:NCH}
\end{align}
Here $P_{L,R}=(1\mp\gamma_5)/2$, $\ell=e,\mu,\tau$
labels charged leptons, and $i$ labels neutrino flavours;
flavour sums are understood. The coefficients $a_{L,R}^\ell$
and $n_{L,R}^i$ have mass dimension $-2$.
Charged-lepton vector terms
are omitted here because their longitudinal projection does
not generate $\bar e i\gamma_5 e$; they remain relevant to other kaon modes.
Let
\begin{equation}
 \begin{gathered}
 \mathcal N=\frac{G_F\alpha_{\rm em}(m_Z)}{\sqrt2\pi s_W^2},\\
 \mathcal Y_\ell=\frac{a_L^\ell-a_R^\ell}{\mathcal N},\qquad
 \mathcal X_i=\frac{n_L^i+n_R^i}{2\mathcal N}.
 \end{gathered}
 \label{S-eq:Ynorm}
\end{equation}
In the standard convention~\cite{Buchalla:1998ba}, $\ham\supset-\frac{G_F}{\sqrt2}\frac{\alpha_{\rm em}}{2\pi s_W^2}\lambda_tY(x_t)(\bar sd)_{V-A}(\bar\ell\ell)_{V-A}$, so that $a_L^{\ell,\rm SM}=+\mathcal N\lambda_tY(x_t)$ and $a_R^{\ell,\rm SM}=0$. Its sign relative to the neutrino coefficient below is fixed by the $Z$ penguin, which couples to $\nu_L$ and $\ell_L$ with opposite weak isospin, and agrees with Ref.~\cite{DAmbrosio:2002vsn}; Eq.~\eqref{S-eq:PEWop} has the same sign, with $a_L^{e,\rm SM}=-2\mathcal P_{\rm EW}$.
The usual neutrino Hamiltonian gives
$n_L^{i,\rm SM}=2\mathcal N\lambda_tX(x_t)$, $n_R^{i,\rm SM}=0$.
These observations fix the normalization of the shifts used in
Eq.~\eqref{S-eq:apparent-plane-shifts}. For a CKM-aligned interaction,
\begin{equation}
 \mathcal Y_\ell=\lambda_t(Y(x_t)+\delta Y_\ell),\qquad
 \mathcal X_i=\lambda_t(X(x_t)+\delta X_i).
 \label{S-eq:MFV}
\end{equation}
For diagonal neutrino couplings the branching ratio is
\begin{equation}
 \BR_L=\frac{\kappa_L}{3\lambda^{10}}
               \sum_{i=e,\mu,\tau}(\Ima\mathcal X_i)^2.
 \label{S-eq:NPnu}
\end{equation}
Real shifts contain no CP phase beyond CKM and yield
Eq.~\eqref{S-eq:apparent-plane-shifts}.

Parity explains the two quark combinations.
Vacuum-to-kaon matrix elements of the vector current vanish,
so $K\to\ell\ell$ selects $a_L-a_R$.
Kaon-to-pion matrix elements of the axial current vanish,
so $K\to\pi\nu\bar\nu$ selects $n_L+n_R$.
Right-handed quark currents therefore enter the charged-lepton and neutrino amplitudes with
opposite relative signs.
For electrons, the kaon-pole contribution to $\Cs^N$ involves the same combination, $a_L^e-a_R^e$.
We assume that the non-pole baryon contributions, which add $30$--$40\%$ in the SM~\cite{Ema:2022yra},
scale in the same way; this holds for left-handed quark currents, as in the SM and for the $Z'$,
whereas right-handed currents would require a separate matching.

The invariant CP projection for an electron coefficient is
$\Ima(\mathcal Y_e\lambda_u^*)$ when the baryon source has
the leading SM weak phase. Eq.~\eqref{eq:plane-CP-probes} uses
the $\lambda_u$-real convention. If new physics also modifies
the nonleptonic nucleon source, its coefficient and phase
must be included using Eq.~\eqref{S-eq:invariant}.

For clarity define $c_i$ as coefficients in the new-physics Hamiltonian
$\ham_{\rm NP}=\sum_i c_iQ_i+\mathrm{h.c.}$.
Thus they are the negative of coefficients defined in a
Lagrangian with the same operator signs. The relevant
Warsaw operators~\cite{Grzadkowski:2010es} are
\begin{align}
 Q_{\ell q}^{(1)}&=(\bar L_\ell\gamma_\mu L_\ell)
                         (\bar q_s\gamma^\mu q_d),\nonumber\\
 Q_{\ell q}^{(3)}&=(\bar L_\ell\gamma_\mu\tau^I L_\ell)
                         (\bar q_s\gamma^\mu\tau^I q_d),\nonumber\\
 Q_{qe}&=(\bar q_s\gamma_\mu q_d)
                         (\bar\ell_R\gamma^\mu\ell_R),\nonumber\\
 Q_{\ell d}&=(\bar L_\ell\gamma_\mu L_\ell)
                         (\bar s_R\gamma^\mu d_R),\nonumber\\
 Q_{ed}&=(\bar\ell_R\gamma_\mu\ell_R)
                         (\bar s_R\gamma^\mu d_R).
 \label{S-eq:Warsaw}
\end{align}
Here $L_\ell=(\nu_{\ell L},\ell_L)^T$. The quark doublets
$q_s,q_d$ have lower components $s_L,d_L$ in the down-quark
mass basis and CKM-rotated upper components;
$\tau^I$ are the Pauli matrices, summed over $I$.
Their tree-level down-sector components are
\begin{align}
 \delta a_L^\ell&=\tfrac12(c_{qe}^\ell
                         -c_{\ell q}^{(1)\ell}
                         -c_{\ell q}^{(3)\ell}),\nonumber\\
 \delta a_R^\ell&=\tfrac12(c_{ed}^\ell-c_{\ell d}^\ell),\nonumber\\
 \delta n_L^\ell&=c_{\ell q}^{(1)\ell}
                         -c_{\ell q}^{(3)\ell},\qquad
 \delta n_R^\ell=c_{\ell d}^\ell .
 \label{S-eq:Warsawmatch}
\end{align}

\Needspace{5\baselineskip}
\subsection{Apparent CKM determinations and additional phases}

To interpret a measurement on the kaon plane, one may solve the
SM expression for its imaginary coordinate while fixing the other
inputs. We call the result the apparent coordinate of that observable: it equals
$A^2\hat\eta$ in the SM, but absorbs any new-physics shift of its lepton channel. The apparent
coordinates of different observables therefore need not agree in the presence of new interactions.
Keeping the SM weak nucleon source and
neglecting the small photon term in these diagnostic expressions,
real CKM-aligned shifts give, at leading Wolfenstein order,
\begin{align}
 \left.(A^2\hat\eta)\right|_{e,\rm SM\ extraction}
   &\simeq y\left[1+\frac{\delta Y_e}{Y(x_t)}\right],\nonumber\\
 \left.|A^2\hat\eta|\right|_{\mu,\rm SM\ extraction}
   &\simeq |y|\left|1+\frac{\delta Y_\mu}{Y(x_t)}\right|,\nonumber\\
 \left.|A^2\hat\eta|\right|_{\nu,\rm SM\ extraction}
   &\simeq |y|\sqrt{\frac13\sum_i
             \left|1+\frac{\delta X_i}{X(x_t)}\right|^2}.
 \label{S-eq:apparent-plane-shifts}
\end{align}
The electron line is written for a signed extraction; a magnitude
measurement takes its absolute value. Muon and neutrino rates
retain discrete sign branches. Unequal real shifts separate the
three determinations without any additional weak phase.
A common fractional shift, $\delta Y_e/Y(x_t)=\delta Y_\mu/Y(x_t)=\delta X_i/X(x_t)$, instead
moves them together; charged-kaon decays, kaon mixing and comparison with $B$ physics test this
direction.
For the exact coordinate definition, the same relations apply to
$\Ima\lambda_t=\lambda_u\lambda^4y/D$ before solving for the
apparent $y$ at a given $x$. The exact relation in
Eq.~\eqref{S-eq:plane-ReIm} retains the corresponding normalization
correction to the leading Wolfenstein expression.

For a complex $\delta Y_\ell$, the signed LO
interpretation instead gives
\begin{equation}
 \left.(A^2\hat\eta)\right|_{\ell,\rm SM\ extraction}
 \simeq y\left[1+\frac{\Rea\delta Y_\ell}{Y(x_t)}\right]
       -x\,\frac{\Ima\delta Y_\ell}{Y(x_t)}.
 \label{S-eq:apparent-plane-phase}
\end{equation}
A new weak phase therefore mixes the real and imaginary quark
amplitudes. In the exact conversion, replace the coefficient $x$
in the CP projection by $x+\lambda^4(x^2+y^2)$ and retain the common
prefactor $\lambda_u\lambda^4/D$. For neutrinos the CP projection
is taken separately for each flavour before summing its square.
The inferred bands need not remain horizontal once these new
coefficients are fitted directly.

\Needspace{5\baselineskip}
\subsection{Direct electron-kaon bound and its EDM interpretation}

The spin-singlet decay coupling satisfies
$|g_{S0}^e|=|g_S^e(m_K^2)|$ in Eq.~\eqref{eq:vertices-main}.
We assume $g_S^e(0)\simeq g_S^e(m_K^2)$, retain the SM weak
nucleon coupling of Sec.~\ref{S-app:matching}, and separate
any electron-EDM contribution.

For a constant pseudoscalar $K_S$ electron coupling, the width
$\Gamma(K_S\to e^+e^-)=m_K\beta_e|g_{S0}^e|^2/(8\pi)$, with
$\beta_e=\sqrt{1-4m_e^2/m_K^2}$, gives
\begin{equation}
 |g_{S0}^e|<
 \left[
 \frac{8\pi\hbar\,\BR(K_S\to ee)_{\rm lim}}
           {\tau_Sm_K\beta_e}
 \right]^{1/2}.
 \label{S-eq:KSbound}
\end{equation}

Using $\tau_S=8.954\times10^{-11}\,\mathrm{s}$,
$\hbar=6.582\times10^{-25}\,\gev\,\mathrm{s}$, and
the KLOE limit $9\times10^{-9}$~\cite{KLOE:2008acb}, one finds
$|g_{S0}^e|<5.78\times10^{-11}$.
The leading SM baryon pole then gives
\begin{equation}
 |\Cs^A|<
 \frac{\sqrt2\,|h_S^A|}{G_Fm_K^2}\,
                     5.78\times10^{-11}.
 \label{S-eq:electronC}
\end{equation}
For thorium and hafnium the resulting coefficients are
$2.28\times10^{-11}$ and $2.27\times10^{-11}$, respectively.
With the ThO conversion factor quoted in the main text and
$\zeta_{\rm HfF^+}=9.17\times10^{-21}\,\ecm$
\cite{Mulder:2025esr}, these translate into upper bounds
on the kaon-mediated equivalent EDM of $3.42\times10^{-31}\,\ecm$ for ThO and
$2.08\times10^{-31}\,\ecm$ for HfF$^+$.
The direct ThO and HfF$^+$ limits,
$1.1\times10^{-29}\,\ecm$ and
$4.1\times10^{-30}\,\ecm$~\cite{ACME:2018yjb,Roussy:2022cmp}, are weaker
by factors of about $32$ and $20$.
These kaon bounds apply only to a purely axial or pseudoscalar, maximally CP-violating $\bar sd$
coupling. A CP-conserving component would also contribute to $K_L\to e^+e^-$, whose analysis must
retain the long-distance dispersive amplitude~\cite{DAmbrosio:2023irq,Hoferichter:2023wiy}. A current with a vector component is limited far more strongly by $K_L\to\pi^0e^+e^-$, whose SM rate
is $3.54\times10^{-11}$ for constructive interference between the indirect and direct CP-violating
amplitudes~\cite{Mescia:2006jd}. Its direct part, $10^{-12}\times4.62(w_V^2+w_A^2)$, depends on the
CP-violating vector- and axial-electron coefficients $w_{V,A}$ defined above Eq.~\eqref{S-eq:LQpimumu},
whose SM values are $0.74$ and $-0.66$. The KTeV bound
$\BR(K_L\to\pi^0e^+e^-)<2.8\times10^{-10}$~\cite{KTeV:2004piee} then allows $|w_A|$, and hence the
kaon-mediated EDM, up to about $11$ times its SM value for fixed $w_V$, or about $10$ times for equal
shifts of $w_V$ and $w_A$, as for the $Z'$.

\Needspace{5\baselineskip}
\subsection{Neutral vector: matching, benchmark, and correlated constraints}

Consider a massive neutral vector coupled to the Hermitian current
\begin{align}
 J^\mu={}&\Delta_L\,\bar s\gamma^\mu P_Ld
             +\Delta_R\,\bar s\gamma^\mu P_Rd+\mathrm{h.c.}
 \nonumber\\
 &+\sum_\ell\left(
 g_L^\ell\bar L_\ell\gamma^\mu L_\ell+
 g_R^\ell\bar\ell_R\gamma^\mu\ell_R\right)+\cdots .
\end{align}
Here $\Delta_{L,R}$ are the complex flavour-changing
quark couplings, while $g_{L,R}^\ell$ are real diagonal lepton
couplings. For $\lag\supset M_{Z'}^2Z'^2/2+Z'_\mu J^\mu$,
integrating out the vector gives
$\ham_{\rm eff}=J_\mu J^\mu/(2M_{Z'}^2)$.
With $g_A^\ell=(g_R^\ell-g_L^\ell)/2$,
\begin{align}
 \delta a_{L,R}^\ell&=
          \frac{\Delta_{L,R}g_A^\ell}{M_{Z'}^2},&
 \delta n_{L,R}^i&=
          \frac{\Delta_{L,R}g_L^i}{M_{Z'}^2},\nonumber\\
 \delta\mathcal Y_\ell&=
  \frac{(\Delta_L-\Delta_R)g_A^\ell}{\mathcal NM_{Z'}^2},&
 \delta\mathcal X_i&=
  \frac{(\Delta_L+\Delta_R)g_L^i}{2\mathcal NM_{Z'}^2}.
 \label{S-eq:Zexact}
\end{align}
Gauge invariance has identified $g_L^{\nu_i}=g_L^i$.

A quark-MFV realization uses
$g_Q\bar Q_L(Y_uY_u^\dagger)\gamma^\mu Q_L$ with real $g_Q$.
Here $Y_u$ is the up-quark Yukawa matrix and $y_t$ its top-quark
eigenvalue.
In the down-quark basis,
$\Delta_L\simeq g_Qy_t^2\lambda_t$.
Taking real lepton couplings and $\Delta_R=0$ gives
\begin{equation}
 \delta Y_\ell=+\frac{g_Qy_t^2g_A^\ell}{\mathcal NM_{Z'}^2},
 \qquad
 \delta X_i=\frac{g_Qy_t^2g_L^i}{2\mathcal NM_{Z'}^2},
 \label{S-eq:MFVZ}
\end{equation}
which are real. Thus an electron-specific $g_R^e$ changes
the molecular EDM branch while the two kaon branches used in
Eq.~\eqref{eq:plane-CP-probes} remain unchanged at tree level.
For a pure left-handed lepton coupling,
$g_A^i=-g_L^i/2$ and $\delta Y_i=-\delta X_i$: the charged-lepton and neutrino amplitudes shift in opposite directions, as for the operator $\mathcal O_{\ell1}$ of Eq.~\eqref{S-eq:MFVops}. Their fractional effects also differ because $Y(x_t)\ne X(x_t)$, and the neutrino rate sums three flavours.
Quark MFV does not impose lepton universality.

For the numerical example in the main text,
we write $g_q=g_Qy_t^2$,
so that $g_q$ is also the leading top coupling. Setting
$g_R^e=g_e$ and all other lepton couplings to zero gives
\begin{equation}
 \delta Y_e=+\frac{g_qg_e}{2\mathcal N M_{Z'}^2},\qquad
 \delta Y_\mu=\delta X_i=0.
 \label{S-eq:Zbenchmark-exact}
\end{equation}
Here $\mathcal N=8.87\times10^{-8}\,\gev^{-2}$ with the
short-distance inputs of Sec.~\ref{S-app:kaonplane}.
The benchmark gives $\delta Y_e=+0.113$ and $\delta Y_e/Y(x_t)=+0.119$. With the EW normalization for thorium this is $|\Delta\Cs^{\rm Th}|\simeq8.2\times10^{-17}$, or $|\Delta d_{e,\rm ThO}^{\rm eq}|\simeq1.2\times10^{-36}\,\ecm$, with the sign of the SM values, which it increases in magnitude; $g_e<0$ reduces them.
Resolving this shift requires improving the present $30\%$
weak-nucleon normalization uncertainty as well as the experimental
precision.

The same exchange produces
\begin{align}
 \ham_{\Delta S=2}^{Z'}&=
 \frac{g_q^2\lambda_t^2}{2M_{Z'}^2}
       (\bar s\gamma_\alpha P_Ld)^2+\mathrm{h.c.},\nonumber\\
 \lag_{eeee}^{Z'}&=-\frac{g_e^2}{2M_{Z'}^2}
       (\bar e_R\gamma_\alpha e_R)^2.
 \label{S-eq:Zcorrelated}
\end{align}
Relative to the SM top-box coefficient in the same quark-current
normalization, the mixing shift is
\begin{equation}
 \frac{\Delta M_{12}}{M_{12}^{\rm SM,tt}}=
 \frac{g_q^2}{2M_{Z'}^2}
 \frac{4\pi^2}{G_F^2m_W^2S_0(x_t)}\,U_{LL}(m_W,M_{Z'}).
 \label{S-eq:Zmixing-ratio}
\end{equation}
The leading QCD evolution factor is approximately $U_{LL}=0.88$
between $5\,\mathrm{TeV}$ and $m_W$, and $S_0(x_t)=2.31$.
Thus the benchmark increases the top-box amplitude by $8.6\%$.
Only the top part of $\varepsilon_K$ has this fractional shift:
the charm--top contribution in Eq.~\eqref{S-eq:plane-epsilon}
must also be retained. The same MFV coefficient affects $B_d$
and $B_s$ mixing, so their measurements constrain this mediator
even though they are not inputs to the SM kaon-plane construction.

LEP uses $\lag_{\rm CI}=-2\pi(\bar e_R\gamma_\alpha e_R)^2/
\Lambda_-^2$ for this sign and chirality. Its limit
$\Lambda_->7.9\,\mathrm{TeV}$ implies
$M_{Z'}/|g_e|>7.9/\sqrt{4\pi}=2.23\,\mathrm{TeV}$
\cite{ALEPH:2013dgf}. The benchmark has $\Lambda_-=17.7\,\mathrm{TeV}$.
At fixed $g_q=0.5$, $M_{Z'}=5\,\mathrm{TeV}$ this one constraint
allows $|g_e|<2.24$, or $|\delta Y_e/Y(x_t)|<0.266$.
The related $e_Rb_L$ contact coefficient is
$|g_eg_q|/M_{Z'}^2=0.020\,\mathrm{TeV}^{-2}$, also below the
LEP bounds on this chirality~\cite{ALEPH:2006jhv}.

\Needspace{5\baselineskip}
\subsection{Muon leptoquark: matching and constraints}\label{S-sec:LQ}

For a canonically normalized $\widetilde S_1$ take
\begin{equation}
 \lag\supset\widetilde S_1
   (\kappa_d\overline{d_R^c}\mu_R+
    \kappa_s\overline{s_R^c}\mu_R)+\mathrm{h.c.}
 \label{S-eq:LQlag}
\end{equation}
All other lepton couplings and the diquark couplings are set to
zero; the latter choice can be enforced by baryon number.
This benchmark does not assume quark MFV.
Integrating out this field and Fierzing gives, in our Hamiltonian
convention~\cite{Dorsner:2016wpm},
\begin{equation}
 c_R^\mu=-\frac{\kappa_d\kappa_s^*}{2M_{\widetilde S_1}^2},
 \qquad \delta a_R^\mu=\frac{c_R^\mu}{2}.
 \label{S-eq:LQmatch}
\end{equation}
The second factor of two comes from the axial component of
$\bar\mu_R\gamma_\alpha\mu_R$.
Consequently the $K_S$ amplitude is multiplied by
$1-\Ima c_R^\mu/[2\mathcal N Y(x_t)\Ima\lambda_t]$, since $\delta\mathcal Y_\mu=-\delta a_R^\mu/\mathcal N$.
For $\Ima\lambda_t=1.42\times10^{-4}$ the denominator is
$2.39\times10^{-11}\,\gev^{-2}$ and the SM branching ratio
with our inputs is $1.89\times10^{-13}$.
Equation~\eqref{eq:LQbenchmark-main} follows.
The chosen coefficient, $c_R^\mu=-i\,1.2\times10^{-10}\,\gev^{-2}$, requires
$\kappa_d\kappa_s^*=+i\,0.00216$ at $M_{\widetilde S_1}=3\,\mathrm{TeV}$,
and can be realized with $|\kappa_d|=|\kappa_s|=0.0465$.
It gives an amplitude enhancement of $6.01$ and a rate
$6.84\times10^{-12}$. Adding the SM $P$ wave, about $5.0\times10^{-12}$~\cite{DAmbrosio:2017klp},
gives a total of $1.2\times10^{-11}$, below the LHCb bound of $2.7\times10^{-11}$ at $90\%$
confidence~\cite{LHCb:2026ksmumu}.

The scalar--lepton box gives
\begin{equation}
 \begin{aligned}
 C_{RR}^{\Delta S=2}(M)&=
       \frac{(\kappa_d\kappa_s^*)^2}{128\pi^2M^2},\\
 M_{12}^{\rm NP}&=\frac13C_{RR}^{\Delta S=2}(\mu)
                       f_K^2m_KB_K(\mu),
 \end{aligned}
 \label{S-eq:LQmix}
\end{equation}
where $M=M_{\widetilde S_1}$~\cite{Dorsner:2016wpm}.
Using $B_K(3\,\gev)=0.55$ and a leading QCD evolution factor
$C_{RR}(3\,\gev)/C_{RR}(3\,\mathrm{TeV})\simeq0.75$ gives
$|2\Rea M_{12}^{\rm NP}|/\Delta m_K\simeq3.9\times10^{-4}$.
A phase $\delta\phi$ added to $c_R^\mu$ would give $M_{12}^{\rm NP}$ the phase $2\delta\phi$,
which contributes to $\varepsilon_K$, and a real part $\Rea c_R^\mu$, which shifts the dispersive
$K_L\to\mu^+\mu^-$ amplitude. The purely imaginary benchmark avoids both: $(\kappa_d\kappa_s^*)^2$ is
real, so $\Ima M_{12}^{\rm NP}=0$ in the $\lambda_u$-real convention, and the dispersive amplitude is
unchanged.

The vector component of the same operator also affects
$K_L\to\pi^0\mu^+\mu^-$. To check this constraint, define the
dimensionless coefficients in
\begin{equation}
 \begin{aligned}
 \ham_{\pi\mu\mu}={}&\frac{G_F\alpha_{\rm em}(m_Z)}{\sqrt2}
 (\bar s\gamma_\alpha d)\\
 &\times[\omega_V\bar\mu\gamma^\alpha\mu+
  \omega_A\bar\mu\gamma^\alpha\gamma_5\mu]+\mathrm{h.c.}
 \end{aligned}
\end{equation}
The overall Hamiltonian sign here follows Eq.~\eqref{S-eq:NCH},
so $\omega_A^{\rm SM}=+\lambda_tY(x_t)/(2\pi s_W^2)$.
Let $w_{V,A}=-\Ima\omega_{V,A}/(1.407\times10^{-4})$, where the sign and the denominator follow the convention of Ref.~\cite{Mescia:2006jd}, whose Hamiltonian has the opposite overall sign. The leptoquark shifts both coefficients by
\begin{equation}
 \delta w_V=\delta w_A=-\frac{\Ima c_R^\mu}
 {4(G_F\alpha_{\rm em}(m_Z)/\sqrt2)(1.407\times10^{-4})}.
\end{equation}
In this convention $w_V^{\rm SM}\simeq0.73$ and $w_A^{\rm SM}=-Y(x_t)/(2\pi s_W^2)$. For $c_R^\mu=-i\,1.2\times10^{-10}\,\gev^{-2}$ the shift is $\delta w_{V,A}=+3.31$, and with the small rescaling to the chosen $\Ima\lambda_t$ one obtains $(w_V,w_A)=(4.05,2.65)$.
The central-value rate formula~\cite{Mescia:2006jd} is
\begin{equation}
 \begin{aligned}
 &\BR(K_L\to\pi^0\mu^+\mu^-)=10^{-12}\bigl[1.09(w_V^2+2.32w_A^2)\\
 &\qquad\mathord{}\pm2.63w_V|a_S|+3.36|a_S|^2+5.2\bigr],
 \end{aligned}
 \label{S-eq:LQpimumu}
\end{equation}
where $a_S$ is the measured $K_S\to\pi^0\ell^+\ell^-$ form-factor
constant.
For $|a_S|=1.2$, the two interference signs give
$(0.33,0.58)\times10^{-10}$, below the $90\%$-confidence limit
$3.8\times10^{-10}$~\cite{KTEV:2000ngj}. Even at the edges
allowed by the $K_S\to\mu^+\mu^-$ limit, $\Ima c_R^\mu\simeq-2.6\times10^{-10}$ and
$+3.1\times10^{-10}\,\gev^{-2}$, the rate stays below this limit, so the $K_S$ dimuon limit is at
present the stronger constraint.

Direct leptoquark searches also leave this particular mass and
small-coupling example open. The recent ATLAS resonant search
reaches $4.3\,\mathrm{TeV}$ for a strange--muon coupling of $3.5$,
but its quoted strange--muon scan starts at coupling $0.5$
\cite{ATLAS:2025upm}; it does not exclude a $3\,\mathrm{TeV}$
state with couplings $0.046$. Pair-production limits are below
this mass.

\Needspace{5\baselineskip}
\subsection{Reach on operators with MFV}
\label{S-sec:MFV}
In MFV the Yukawa matrices are the only sources of flavour
breaking~\cite{DAmbrosio:2002vsn}. The $s\to d$ component of a left-handed quark current
then carries $(Y_uY_u^\dagger)_{sd}\simeq y_t^2\lambda_t$ in the down-quark mass basis. In the
notation of Eq.~\eqref{S-eq:Warsaw} we consider
\begin{equation}
 \mathcal O_{\ell1}=\sum_{\ell=e,\mu,\tau}Q_{\ell q}^{(1)},\qquad
 \mathcal O_e=Q_{qe}\ (\ell=e),
 \label{S-eq:MFVops}
\end{equation}
each with coefficient $c=\pm\lambda_t/\Lambda^2$ in $\ham_{\rm NP}$; $y_t^2\simeq1$ is absorbed
into $\Lambda$.
The operator $\mathcal O_{\ell1}$ is the lepton-universal operator of
Ref.~\cite{DAmbrosio:2002vsn}. The operator $\mathcal O_e$ couples only to right-handed electrons, as
the $Z'$ example does, since quark MFV does not require lepton universality. Projecting the lepton
currents as in Eq.~\eqref{S-eq:Warsawmatch} and defining
\begin{equation}
 \epsilon=\frac{c}{2\mathcal N\lambda_t}=\pm\frac{\Lambda_0^2}{\Lambda^2},\qquad
 \Lambda_0=(2\mathcal N)^{-1/2}=2.37\,\mathrm{TeV},
 \label{S-eq:MFVeps}
\end{equation}
the loop functions of Eq.~\eqref{S-eq:MFV} are shifted by
\begin{equation}
 \mathcal O_{\ell1}:\ \delta X=+\epsilon,\quad \delta Y_e=\delta Y_\mu=-\epsilon;\qquad
 \mathcal O_e:\ \delta Y_e=+\epsilon .
 \label{S-eq:MFVshifts}
\end{equation}
The vector lepton couplings also shift the CP-violating coefficients of $K_L\to\pi^0\ell^+\ell^-$.
With $w_{V,A}$ defined above Eq.~\eqref{S-eq:LQpimumu}, whose SM values at the benchmark $\Ima\lambda_t$ are
$w_V=0.74$ and $w_A=-0.66$,
\begin{equation}
 \begin{aligned}
 \mathcal O_{\ell1}:&\quad
 \delta w_V=-\delta w_A=-\kappa_w\epsilon,\\
 \mathcal O_e:&\quad
 \delta w_V^e=\delta w_A^e=-\kappa_w\epsilon.
 \end{aligned}
 \label{S-eq:MFVw}
\end{equation}
with $\kappa_w=\Ima\lambda_t/(2\pi s_W^2\times1.407\times10^{-4})=0.695$.

Relative to the SM, the observables then scale as
\begin{align}
 \frac{\Cs^A}{\Cs^{A,\rm SM}}&=1+\frac{\delta Y_e}{Y},\qquad
 \frac{\BR_S^{(0)}}{\BR_S^{(0),\rm SM}}=\Big(1+\frac{\delta Y_\mu}{Y}\Big)^2,\nonumber\\
 \frac{\BR_L}{\BR_L^{\rm SM}}&=\Big(1+\frac{\delta X}{X}\Big)^2,
 \label{S-eq:MFVratios}
\end{align}
with $X\equiv X(x_t)$ and $Y\equiv Y(x_t)$. The charged mode follows from Eq.~\eqref{S-eq:plane-Kplus}
with $X\to X+\delta X$, and $K_L\to\ell^+\ell^-$ from Eq.~\eqref{S-eq:plane-KLmu} with
$\chi_{\rm short}=-\widetilde\kappa[\Rea\lambda_t(Y+\delta Y_\ell)+\Rea\lambda_cY_{\rm NL}]$, the same for
both leptons because the lepton mass cancels in the diphoton normalization. We take
$\Ima\mathcal A_{\rm LD}$ from unitarity ($-5.21$ for muons and $-21.6$ for electrons) and fix the
dispersive parts to central values: $\Rea\mathcal A_{\rm LD}^{\mu\mu}=-0.66$, one of the two solutions of
Sec.~\ref{S-app:kaonplane}, for illustration, and $\Rea\mathcal A_{\rm LD}^{ee}=28.3$ from the VMD model of Sec.~\ref{S-app:loop}
continued to $t=m_K^2$, with the source sign chosen there. The short-distance term then changes
$\BR(K_L\to e^+e^-)$ by only $8\%$. The rate of $K_L\to\pi^0\mu^+\mu^-$ follows from
Eq.~\eqref{S-eq:LQpimumu} and the electron mode from~\cite{Mescia:2006jd}
\begin{equation}
 \begin{aligned}
 \BR(K_L\to\pi^0e^+e^-)={}&10^{-12}\big[14.5|a_S|^2+11.3\,w_V|a_S|\\
 &+4.62(w_V^2+w_A^2)\big],
 \end{aligned}
 \label{S-eq:MFVpiee}
\end{equation}
both with constructive interference and $|a_S|=1.2$. Finally, the untagged $K_S\to e^+e^-$ rate is
dominated by the CP-conserving two-photon $P$ wave,
$\BR\simeq2\times10^{-14}$~\cite{Ecker:1991ru,KLOE:2008acb}. The short-distance $S$ wave,
\begin{equation}
 \BR_S^{(0)}(ee)=\BR_S^{(0)}(\mu\mu)\,\frac{m_e^2\beta_e}{m_\mu^2\beta_\mu}=4.9\times10^{-18},
 \label{S-eq:MFVKSee}
\end{equation}
is a fraction $2.4\times10^{-4}$ of the total and scales as $(1+\delta Y_e/Y)^2$.

\begin{figure*}[!t]
 \centering
 \includegraphics[width=\textwidth]{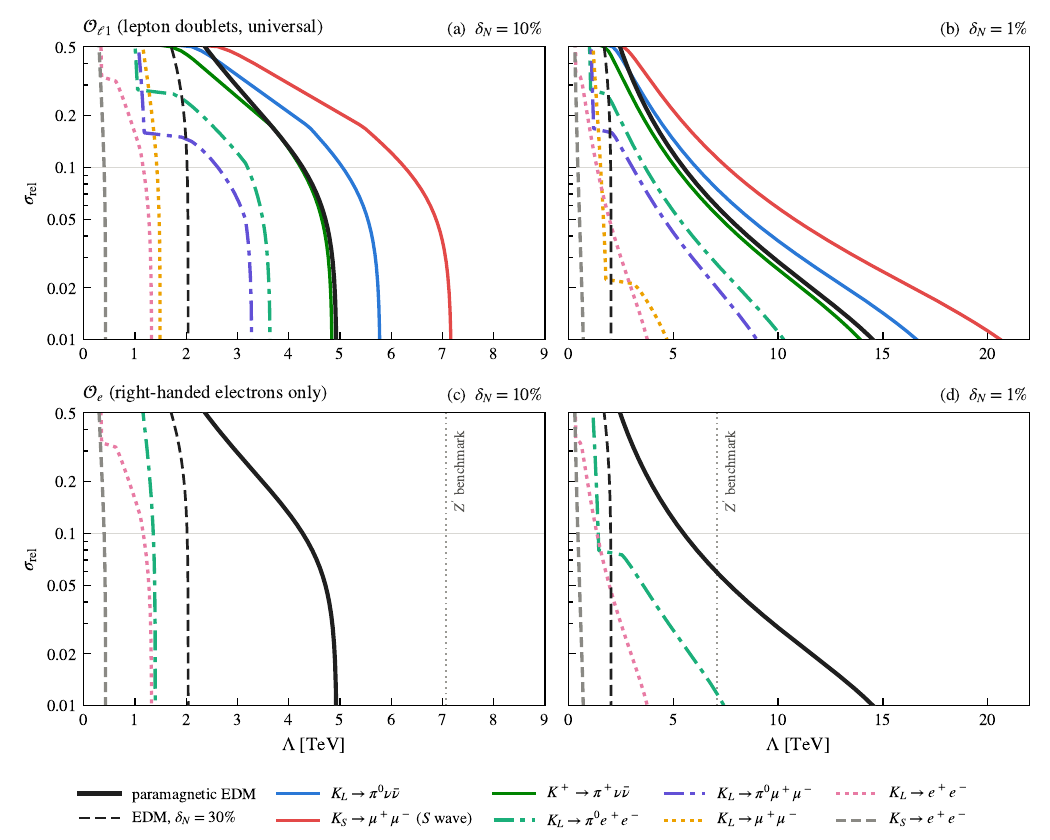}
 \caption{Reach on the scale $\Lambda$ of the MFV operators of
 Eq.~\eqref{S-eq:MFVops} for a hypothetical measurement with relative precision
 $\sigma_{\rm rel}$ and central value equal to the SM prediction, as in Fig.~4 of
 Ref.~\cite{DAmbrosio:2002vsn}; $\sigma_{\rm rel}$ refers to the rate for the decays and to the
 equivalent EDM for molecules. Top: lepton-universal $\mathcal O_{\ell1}$. Bottom: $\mathcal O_e$,
 coupled only to right-handed electrons. Left (right) panels assume a $10\%$ ($1\%$) uncertainty
 $\delta_N$ of the SM prediction of each observable; the black dashed curve uses the present $30\%$
 weak-nucleon uncertainty for the EDM. Each curve is the $95\%$ confidence reach of
 Eq.~\eqref{S-eq:MFVcrit} for the weaker sign of $\epsilon$. The vertical grey dotted line marks the $Z'$
 benchmark, $\Lambda=7.1\,$TeV.}
 \label{S-fig:MFV}
\end{figure*}

Following Ref.~\cite{DAmbrosio:2002vsn}, Fig.~\ref{S-fig:MFV} shows the scale $\Lambda$ that a
measurement with relative precision $\sigma_{\rm rel}$, and central value equal to the SM
prediction, would probe. Here $\sigma_{\rm rel}$ refers to the quantity measured: the rate for the
decays and the equivalent EDM, linear in the amplitude, for molecules. Let $R$ be the observable
divided by its SM value and $\delta_N$ the relative uncertainty of its SM prediction from CKM and
hadronic inputs. Since this normalization multiplies the new-physics term as well, a value of
$\Lambda$ is excluded at $95\%$ confidence when
\begin{equation}
 |R-1|>1.96\,\big(\sigma_{\rm rel}^2+\delta_N^2R^2\big)^{1/2}.
 \label{S-eq:MFVcrit}
\end{equation}
Each curve gives the largest excluded $\Lambda$ for the weaker of the two signs of $\epsilon$. For
destructive interference a rate first decreases and returns to its SM value near $\epsilon\simeq-2F$,
with $F=X$ or $Y$. Where the attainable deficit stays below the bound, only the overshoot at larger
$|\epsilon|$ is excluded, which produces the horizontal segments in Fig.~\ref{S-fig:MFV}.

As $\sigma_{\rm rel}\to0$ the reach saturates at a value set by $\delta_N$. For $\mathcal O_{\ell1}$ and
$\delta_N=10\%$ it is $7.2\,$TeV for the $K_S\to\mu^+\mu^-$ $S$ wave, $5.8\,$TeV for
$K_L\to\pi^0\nu\bar\nu$, $4.9\,$TeV for the EDM, $4.8\,$TeV for $K^+\to\pi^+\nu\bar\nu$, $3.6$ and $3.3\,$TeV for
$K_L\to\pi^0e^+e^-$ and $K_L\to\pi^0\mu^+\mu^-$, and $1.5$ and $1.3\,$TeV for $K_L\to\mu^+\mu^-$ and
$K_L\to e^+e^-$. For $\delta_N=1\%$ they grow by about $\sqrt{10}$ as $\sigma_{\rm rel}\to0$; at
$\sigma_{\rm rel}=1\%$, the lower edge of Fig.~\ref{S-fig:MFV}, they are about $15\%$ smaller,
e.g.\ $14.5\,$TeV for the EDM. The EDM and the dimuon
$S$ wave probe the same amplitude. The factor of about $\sqrt2$ between their reaches arises
because the rate is quadratic in it, so the same $\delta_N$ and $\sigma_{\rm rel}$ on a rate correspond
to half the amplitude precision; at equal amplitude precision the two curves coincide.

For $\mathcal O_e$ only electron observables respond, and the EDM has the largest reach. Next comes
$K_L\to\pi^0e^+e^-$, with $1.4\,$TeV for the weaker sign and $2.9\,$TeV for the other at
$\delta_N=10\%$: for $\epsilon>0$ its interference terms partly cancel and keep the rate within $16\%$ of
the SM until the new amplitude dominates. $K_L\to e^+e^-$ and $K_S\to e^+e^-$ stay below $1.5$ and
$0.5\,$TeV, because their rates are dominated by two-photon exchange. The $Z'$ benchmark of the
Letter corresponds to $\Lambda=M_{Z'}/\sqrt{g_qg_e}=7.1\,$TeV. Its EDM shift, $\delta Y_e/Y=0.12$,
requires $\delta_N$ below about $5\%$. With the present $30\%$ weak-nucleon uncertainty the EDM reach is
$2.0\,$TeV, or $4.0\,$TeV for the sign that increases the signal.

The curves for $K_L\to\ell^+\ell^-$, $K_S\to e^+e^-$ and $K_L\to\pi^0\ell^+\ell^-$ assume the long-distance
inputs above with no further uncertainty. Present uncertainties, for example $\pm0.38$ on
$\Rea\mathcal A_{\rm LD}^{\mu\mu}$ and the second dispersive solution $0.16$, or the uncertainty on
$|a_S|$, exceed $\delta_N=1\%$, so these curves are illustrative.

\Needspace{5\baselineskip}
\subsection{Light mediators and scalar operators}

The analysis above assumes heavy new physics with the vector and axial structure of the SM
operator, so that one local coefficient describes molecules and kaon decays alike. Two departures,
quoted in the Letter, change how the determinations compare.

New physics lighter than the kaon need not be local at the relevant momentum transfers. A light
spin-zero mediator with $\lag\supset-\phi(g_N^S\bar N N+g_e^P\bar e i\gamma_5 e)$ gives
\begin{equation}
 \Delta\Cs^N(-Q^2)=\frac{\sqrt2}{G_F}
                 \frac{g_N^Sg_e^P}{M_\phi^2+Q^2}.
 \label{S-eq:lightkernel}
\end{equation}
Molecules probe $Q\sim Z\alpha m_e$, a few hundred keV. If the mediator also couples to $\bar sd$,
kaon decays probe its propagator at $t=m_K^2$ instead, so for $M_\phi\lesssim m_K$ the two
determinations no longer share a single coefficient. For $M_\phi\lesssim Q$ the interaction is not
local even in the molecule: the Yukawa potential must be integrated over the nuclear density with
molecular electronic wave functions, and the relation between $g_N^Sg_e^P$ and the measured shift
depends on $M_\phi$~\cite{Stadnik:2017hpa}.

Scalar and pseudoscalar $\bar sd\,\mu\mu$ operators lift the helicity suppression and change the
partial-wave content of the dimuon modes. Writing
\begin{equation}
 \mathcal M(K_j\to\ell^+\ell^-)
 =\bar u(g_{j1}^\ell+i g_{j0}^\ell\gamma_5)v,
 \label{S-eq:partialwaves}
\end{equation}
$g_{j0}^\ell$ is the spin-singlet $S$ wave and $g_{j1}^\ell$ the spin-triplet $P$ wave. For the SM
and the vector and axial operators above, the short-distance amplitudes are pure $S$ wave, so
$g_{L1}^\mu=0$. A scalar muon operator such as $(\bar L_\mu\mu_R)(\bar s_Rq_d)$, with weak-isospin
indices contracted, generates both partial waves. If $g_{L1}^\mu\ne0$, the tagged interference also
contains $\beta_\mu^2g_{S1}^\mu(g_{L1}^\mu)^*$, and the unpolarized time dependence alone no longer
isolates the CP-violating $S$ wave~\cite{Dery:2021vql}; the muon determination of
Eq.~\eqref{eq:plane-CP-probes} then requires additional observables.

\end{document}